 \documentclass[final,3p,times]{elsarticle}
\usepackage{multirow,setspace,times,amssymb,amsmath,graphicx,color,rotating,subfigure,url}
\usepackage{lineno}
\usepackage{natbib}

\journal{Physica A} %% change this journal name and put the correct one

\begin{document}

\begin{frontmatter}

\title{Scaling and memory in the return intervals of realized volatility}
\author[SB,RCE,RCSE]{Fei Ren}
\author[SB,RCE,SS]{Gao-Feng Gu}
\author[SB,RCE,RCSE,SS,RCFE]{Wei-Xing Zhou\corref{cor}}
\cortext[cor]{Corresponding author. Address: 130 Meilong Road, P.O.
Box 114, School of Business, East China University of Science and
Technology, Shanghai 200237, China, Phone: +86 21 64253634, Fax: +86
21 64253152.}
\ead{wxzhou@ecust.edu.cn} %

\address[SB]{School of Business, East China University of Science and Technology, Shanghai 200237, China}
\address[RCE]{Research Center for Econophysics, East China University of Science and Technology, Shanghai 200237, China}
\address[RCSE]{Engineering Research Center of Process Systems Engineering (Ministry of Education), East China University of Science and Technology, Shanghai 200237, China}
\address[SS]{School of Science, East China University of Science and Technology, Shanghai 200237, China}
\address[RCFE]{Research Center on Fictitious Economics \& Data Science, Chinese Academy of Sciences, Beijing 100080, China}

\begin{abstract}
We perform return interval analysis of 1-min {\em{realized
volatility}} defined by the sum of absolute high-frequency intraday
returns for the Shanghai Stock Exchange Composite Index (SSEC) and
$22$ constituent stocks of SSEC. The scaling behavior and memory
effect of the return intervals between successive realized
volatilities above a certain threshold $q$ are carefully
investigated. In comparison with the volatility defined by the
closest tick prices to the minute marks, the return interval
distribution for the realized volatility shows a better scaling
behavior since 20 stocks (out of 22 stocks) and the SSEC pass the
Kolmogorov-Smirnov (KS) test and exhibit scaling behaviors, among
which the scaling function for $8$ stocks could be approximated well
by a stretched exponential distribution revealed by the KS
goodness-of-fit test under the significance level of 5\%. The
improved scaling behavior is further confirmed by the relation
between the fitted exponent $\gamma$ and the threshold $q$. In
addition, the similarity of the return interval distributions for
different stocks is also observed for the realized volatility. The
investigation of the conditional probability distribution and the
detrended fluctuation analysis (DFA) show that both short-term and
long-term memory exists in the return intervals of realized
volatility.
\end{abstract}


\end{frontmatter}

\section{Introduction}

The study of extreme events has drawn much attention of scientists,
for instance the nature records of floods, temperatures and
earthquakes
\cite{Bunde-Eichner-Havlin-Kantelhardt-2003-PA,Bunde-Eichner-Havlin-Kantelhardt-2004-PA,Bunde-Eichner-Kantelhardt-Havlin-2005-PRL,Bak-Christensen-Danon-Scanlon-2002-PRL,Corral-2004-PRL,Saichev-Sornette-2006-PRL}.
By investigating the return intervals between successive extreme
events exceeding a certain threshold $q$, scaling behaviors are
revealed in the return interval distributions for numerous complex
systems
\cite{Bunde-Eichner-Kantelhardt-Havlin-2005-PRL,Bak-Christensen-Danon-Scanlon-2002-PRL,Corral-2004-PRL,Saichev-Sornette-2006-PRL}.
As with this scaling behavior, we can infer that the probability
distribution of the return intervals of rare events that are
difficult to take empirical measurements. This scaling property of
extreme events is supposed to be of great importance for the risk
assessment of rare events. Further studies have shown that the
scaling behavior of return intervals may arise from the long-term
memory of the original records
\cite{Pennetta-2006-EPJB,Olla-2007-PRE,Eichner-Kantelhardt-Bunde-Havlin-2006-PRE,Eichner-Kantelhardt-Bunde-Havlin-2007-PRE,Bogachev-Eichner-Bunde-2007-PRL}.
This suggests that the scaling behavior might also appear in other
types of records with long-term correlations, such as the stock
market records.

The early study of extreme events in stock markets mainly concerns
the return intervals between successive volatilities above a certain
threshold $q$. The time intervals between consecutive trades and
orders have also been widely studied
\cite{Mainardi-Raberto-Gorenflo-Scalas-2000-PA,Sabatelli-Keating-Dudley-Richmond-2002-EPJB,Ivanov-Yuen-Podobnik-Lee-2004-PRE,Scalas-Gorenflo-Luckock-Mainardi-Mantelli-Raberto-2004-QF,Scalas-Kaizoji-Kirchler-Huber-Tedeschi-2006-PA,Jiang-Chen-Zhou-2008-PA}.
In general, the volatility $R(t,\delta t)$ is simply defined as the
magnitude of logarithmic return between tick prices at time $t$ and
$t-\delta t$, i.e. $R(t,\delta t)=|\ln(Y(t))-\ln(Y(t-\delta t))|$,
where $Y(t)$ is the tick price at time $t$. With this definition,
Yamasaki et al and Wang et al used the daily data and intraday data
of US stocks to study the probability distribution of volatility
return intervals, and indeed found a scaling behavior
\cite{Yamasaki-Muchnik-Havlin-Bunde-Stanley-2005-PNAS,Wang-Yamasaki-Havlin-Stanley-2006-PRE,Wang-Weber-Yamasaki-Havlin-Stanley-2007-EPJB,VodenskaChitkushev-Wang-Weber-Yamasaki-Havlin-Stanley-2008-EPJB}.
Further studies show that long-term memory also exists in the
volatility return intervals. Similar scaling behavior and long-term
memory are observed in thousands Japanese stocks and 4 Chinese
stocks
\cite{Jung-Wang-Havlin-Kaizoji-Moon-Stanley-2008-EPJB,Qiu-Guo-Chen-2008-PA}.
On the contrary, there is growing evidence showing that the return
interval distribution may exhibit multiscaling behavior. Lee et al
studied the 1-min volatility data of the Korean KOSPI index
\cite{Lee-Lee-Rikvold-2006-JKPS}, and Wang et al analyzed the
trade-by-trade data of $500$ stocks composing the S\&P 500 index and
1137 US common stocks
\cite{Wang-Yamasaki-Havlin-Stanley-2008-PRE,Wang-Yamasaki-Havlin-Stanley-2009-PRE},
and Kaizoji analyzed the daily data of 800 companies listed on Tokyo
Stock Exchange and the Nikkei 225 index
\cite{Kaizoji-Kaizoji-2004b-PA}, and Ren and Zhou studied two
Chinese stock market indexes \cite{Ren-Zhou-2008-EPL}, and all these
studies show that the return interval distributions for different
thresholds $q$ exhibit a systematic deviation from scaling and show
multiscaling behavior. Ren, Guo and Zhou conducted a more careful
study to scrutinize 30 very liquid stocks in the Chinese market
using the Kolmogorov-Smirnov (KS) test, and found that some stocks
pass the KS test displaying scaling behaviors while others show
multiscaling behaviors \cite{Ren-Guo-Zhou-2009-PA}.

In the finance literature, there are many different estimators for
volatility. Anderson et al proposed a daily realized volatility
constructed from the sum of the square intraday returns
\cite{Andersen-Bollerslev-Diebold-Labys-2001-JASA,Andersen-Bollerslev-Diebold-Labys-2001-JFE}.
This realized daily volatility contains more valuable information
about the intraday data, and is made arbitrarily close to the
underlying integrated volatility. In addition, there are other daily
volatility estimators constructed based on the intraday data
\cite{Schwert-1990-FAJ,Hsieh-1991-JF,denHaan-Levin-1996-NBER,Bollen-Inder-2002-JEF}.
Subsequently, this realized volatility is generalized to the
volatility which sums the square returns in a fixed time interval
\cite{Andersen-Bollerslev-Diebold-2009-HFE}. This realized
volatility may better describe the price fluctuation caused by the
trades occurring in that time interval. It is interesting to
investigate the statistical properties of return intervals of
realized volatility.

In this paper, we investigate the statistical properties of the
return intervals of realized volatility based on high-frequency
intraday data in the Chinese stock market. Inspired by the sum of
squared returns realized volatility raised by Anderson, we introduce
an estimate of 1-min realized volatility by summing the absolute
logarithmic returns utilizing all trading data in each minute. This
1-min realized volatility contains more valuable information which
lies in the high frequency trading data. Using the
Kolmogorov-Smirnov (KS) test and detrended fluctuation analysis
(DFA) method, we test if the scaling behavior and long-term memory
of the return intervals maintain with this 1-min realized
volatility.

The paper proceeds as follows: In Section \ref{S1:VolDef}, we
introduce the database analyzed and the definition of 1-min realized
volatility. In Section \ref{S1:PDF}, we study the return interval
distribution of the realized volatility using the KS tests. In
Section \ref{S1:Memory}, we further study the memory effect of the
realized volatility return intervals. Section \ref{S1:Conclusion}
summarizes.

\section{Volatility definition}
\label{S1:VolDef}

Our analysis is based on the high-frequency intraday data of SSEC
and $22$ liquid stocks traded on the Shanghai Stock Exchange. These
$22$ stocks are the most actively traded stocks representative of a
variety of industry sectors, and consequently have the largest sizes
among all the stocks. The prices and the associated times of the
SSEC index and the individual stocks are recorded every six to eight
seconds from January 2004 to June 2006.

In many previous studies, the volatility is defined as the magnitude
of the logarithmic return,
\begin{equation} \label{Eq:volatility1}
R_1(t)=|\ln Y(t)-\ln Y(t-1)|,
\end{equation}
where Y(t) is the closest tick price to a minute mark $t$
\cite{Wang-Yamasaki-Havlin-Stanley-2006-PRE,Wang-Weber-Yamasaki-Havlin-Stanley-2007-EPJB,VodenskaChitkushev-Wang-Weber-Yamasaki-Havlin-Stanley-2008-EPJB,Lee-Lee-Rikvold-2006-JKPS,Ren-Zhou-2008-EPL,Ren-Guo-Zhou-2009-PA}.
In this paper, we focus on the realized volatility constructed from
the sum of absolute trade-by-trade returns within one minute.
Suppose that $Y(t')$ is the tick price at time $t'$, then the
realized volatility is defined as
\begin{equation} \label{Eq:volatility2}
R_2(t)=\sum_{t-1<t'\leqslant{t}}|\ln Y(t')-\ln Y(t'-1)|,
\end{equation}
where the sum is taken over all the tick times between $t-1$ and
$t$. For both volatility definitions the sampling time is one
minute, and the volatility data size is about 140,000 for each
stock.

Before doing the analysis, we removed the intraday pattern to
eliminate its periodic effect on the return interval distribution
\cite{Ni-Zhou-2009-JKPS,Ren-Guo-Zhou-2009-PA} via dividing the
volatility $R_i(t)$ by its average value corresponding to time $t$
on the day
\cite{Wang-Yamasaki-Havlin-Stanley-2006-PRE,Wang-Weber-Yamasaki-Havlin-Stanley-2007-EPJB,VodenskaChitkushev-Wang-Weber-Yamasaki-Havlin-Stanley-2008-EPJB}.
Then we normalize the data by dividing its standard deviation so
that the volatility is in units of its standard deviation.

\section{Probability distribution of realized volatility return intervals}
\label{S1:PDF}

\subsection{Empirical return interval distribution}

We study the return intervals $\tau$ between successive volatilities
exceeding a certain threshold $q$. For each threshold $q$, a series
of return intervals is obtained  and its empirical probability
distribution $P_q(\tau)$ can be obtained. Many empirical studies
have showed that the probability distribution function (PDF) of the
scaled return intervals may obey a scaling form
\begin{equation}
P_q(\tau)=\frac{1}{\langle \tau \rangle} f ( \tau / \langle \tau
\rangle ), \label{Eq:pdf:scaling}
\end{equation}
where $\langle \tau \rangle$ is the mean return interval which
depends on the threshold $q$.

\begin{figure}[h]
\centering
\includegraphics[width=6cm]{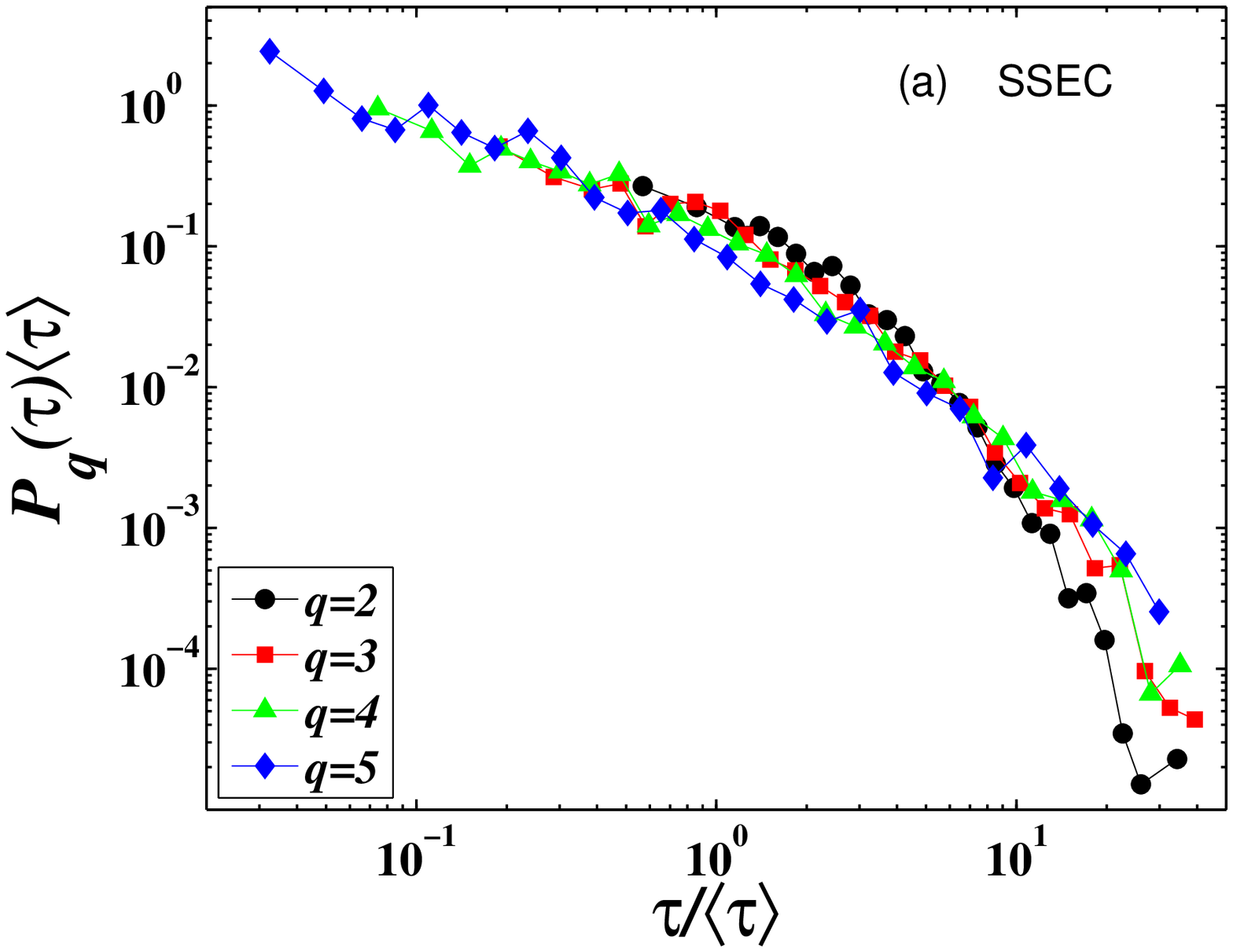}
\includegraphics[width=6cm]{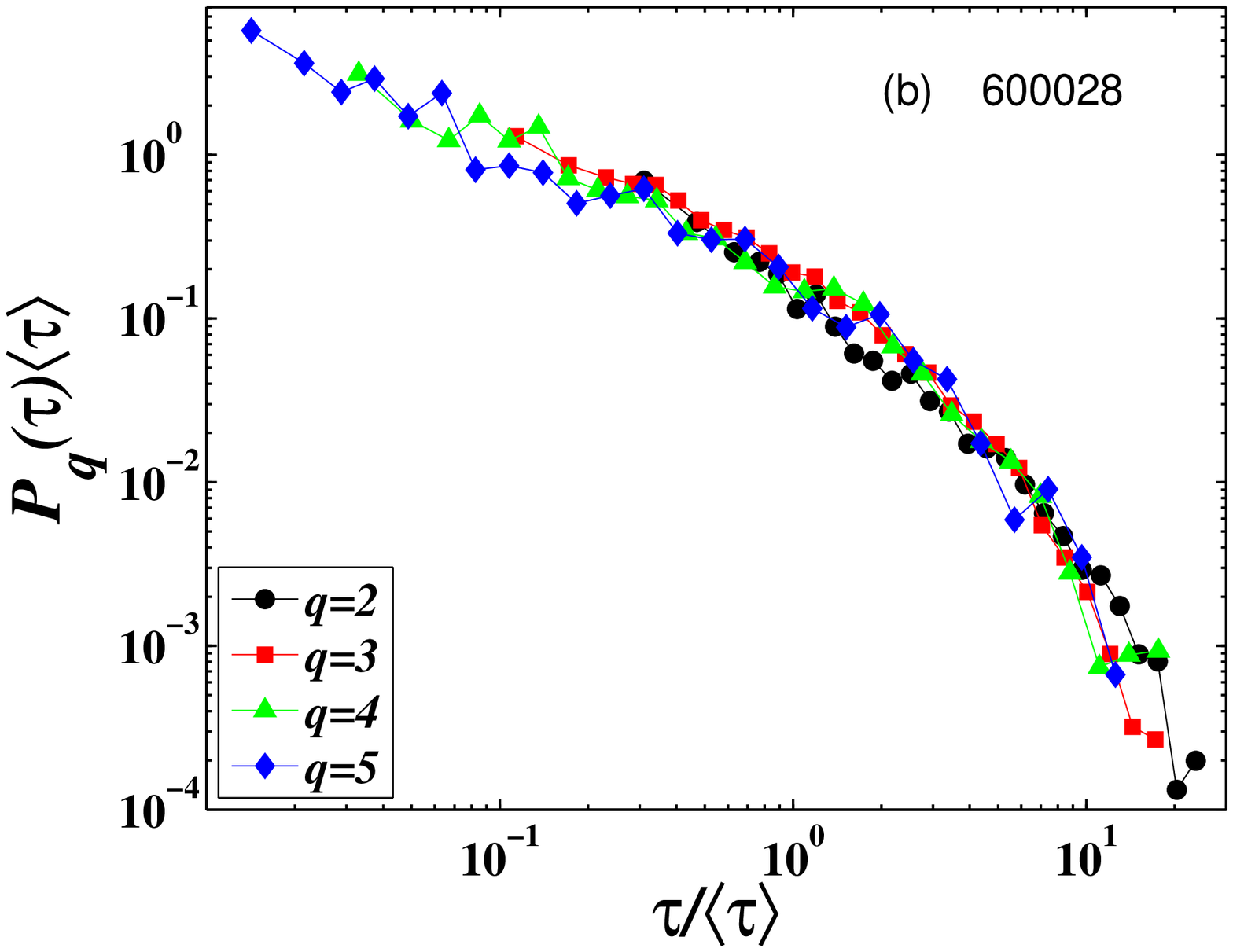}
\includegraphics[width=6cm]{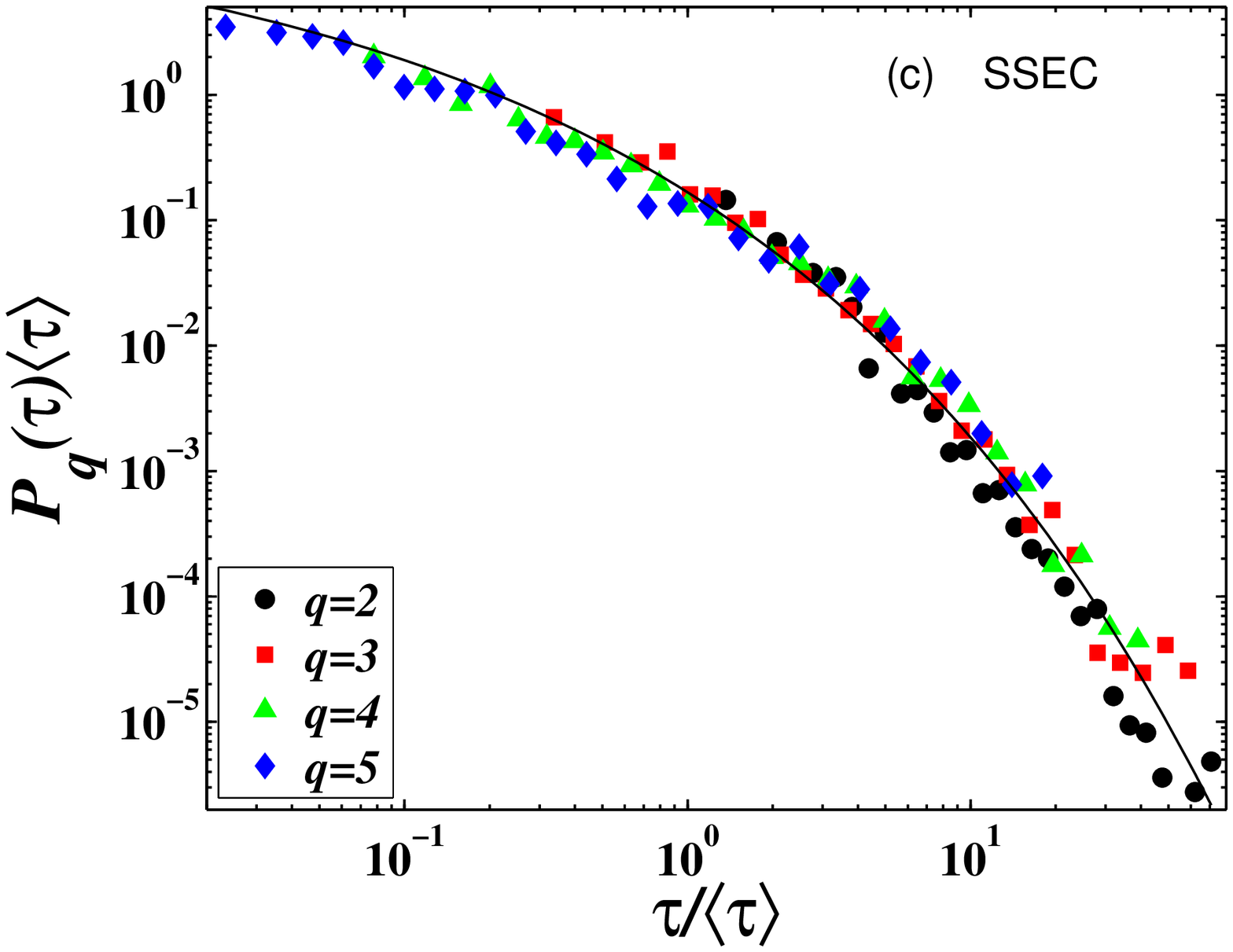}
\includegraphics[width=6cm]{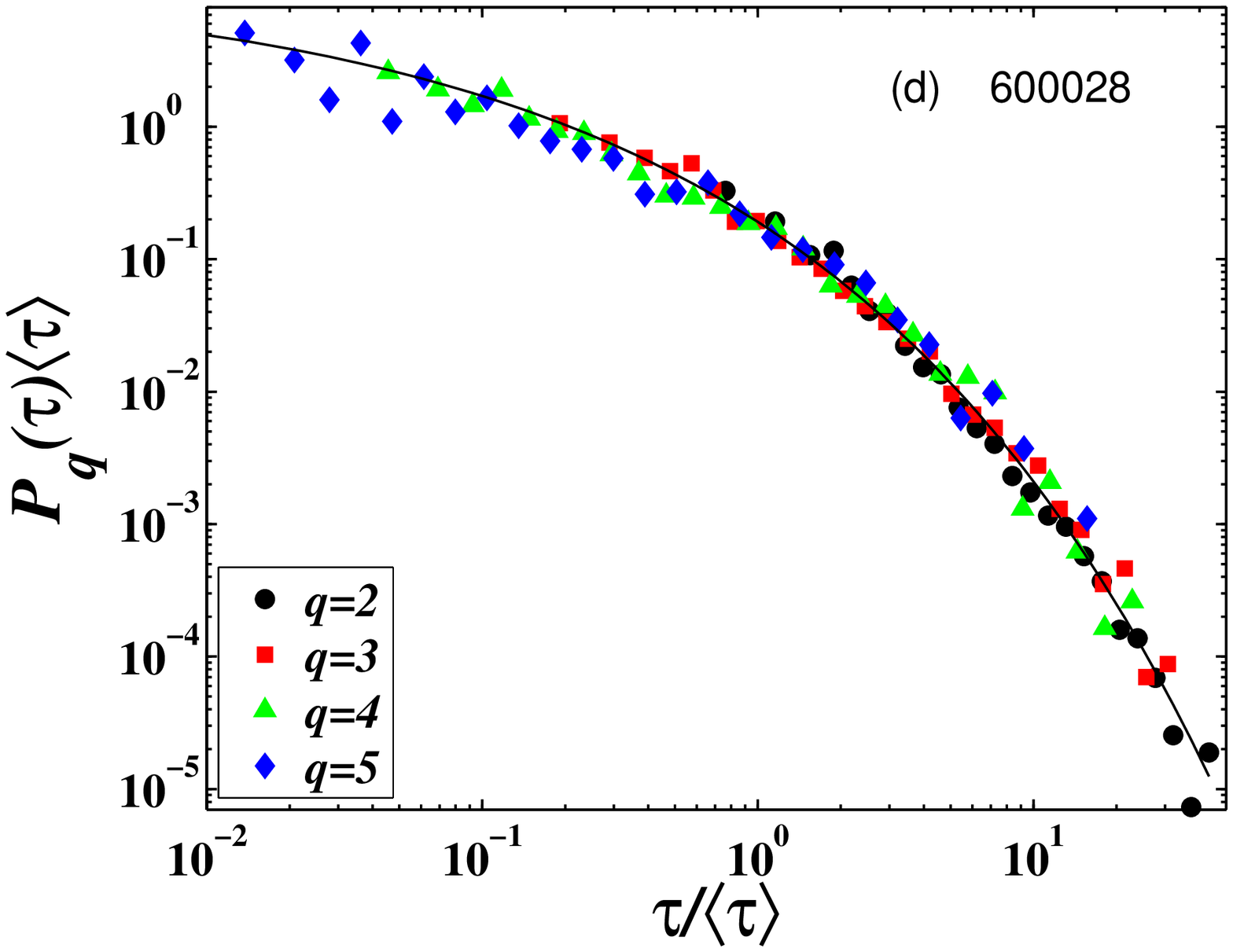}
\caption{(Color online) Probability distributions of the scaled
return intervals of volatilities defined by (a) $R_1$ for SSEC, (b)
$R_1$ for stock $600028$, (c) $R_2$ for SSEC, and (d) $R_2$ for
stock $600028$. Solid curves in (c) and (d) are stretched
exponential fits with $\gamma=0.26$ and $0.31$ respectively.}
\label{Fig:RI:PDF}
\end{figure}

Ren et al have studied the return intervals of the volatility
defined by $R_1$ for the SSEC and the $22$ constituent stocks
analyzed in this paper
\cite{Ren-Zhou-2008-EPL,Ren-Guo-Zhou-2009-PA}. They found that, for
some of the stocks and the SSEC, the return interval distributions
do not show scaling behavior. In Fig. \ref{Fig:RI:PDF} (a) and (b),
the PDFs of the scaled return intervals of volatility $R_1$ for the
SSEC and a representative stock 600028 are plotted. For the SSEC,
the curves for different thresholds $q=2,3,4,5$ do not collapse to a
single curve and show systematic deviation from scaling. For stock
600028, the deviation becomes relatively small, but one can still
see some difference between the curves. Fig. \ref{Fig:RI:PDF} (c)
and (d) plot the return interval PDFs of the realized volatility
$R_2$ for the SSEC and stock $600028$ for comparison. We find that
the curves for different $q$ values approximately collapse to a
single solid curve. This strongly suggests that the return interval
distribution for the realized volatility shows better scaling
behavior than that of the volatility defined by $R_1$.

\subsection{Kolmogorov-Smirnov test of scaling in $P_q(\tau)$ for different $q$}

Eyeballing of the return interval distributions offers a qualitative
way of distinguishing the scaling and nonscaling behavior. A
quantitative method, the Kolmogorov-Smirnov (KS) test, is further
adopted to examine the possible collapse of the interval
distributions for different thresholds. We use the KS test to
examine two return interval distributions for $q=2$ and $5$, which
seems to behave most differently among all the distributions. If the
distributions for $q=2$ and $5$ pass the test, we can conclude that
all the return interval distributions for different $q$ values
collapse onto a single curve and consequently obey a scaling law.
Suppose that $F_2$ is the cumulative distribution function (CDF) of
return intervals for $q=2$ and $F_5$ is the CDF of return intervals
for $q=5$. We calculate the $KS$ statistic by comparing the two CDFs
in the overlapping region:
\begin{equation}
   KS = \max\left(|F_2-F_5|\right)~.
   \label{Eq:KS}
\end{equation}
When the $KS$ statistic is smaller than a critical value denoted by
$CV$ (i.e., $KS<CV$), we can conclude that the distribution for
$q=2$ is coincident with the distribution for $q=5$, and the return
interval distributions show scaling behavior. The critical value is
$CV=c_\alpha /\sqrt{{mn}/({m+n})}$, where $m$ and $n$ are the
numbers of interval samples for $q_i$ and $q_j$
\cite{Darling-1957-AMS,Stephens-1974-JASA}, and the threshold is
$c_\alpha=1.36$ at the significance level of $\alpha=5\%$
\cite{Smirnov-1948-AMS,Young-1977-JHcCc}.

Ren, Guo and Zhou have used the KS test to examine the scaling
behavior of the return interval distribution for the volatility
defined by $R_1$, and found that among the $22$ chosen stocks, $11$
of the individual stocks pass the KS test and show scaling behavior,
while the remaining $11$ stocks as well as the SSEC do not show
scaling behavior as depicted in Table \ref{TB:KStest}. We also use
the KS test to study the scaling behavior of the return interval
distribution for the realized volatility defined by $R_2$.
Remarkably, only $2$ stocks fail in the test, and the remaining $20$
stocks pass the test and show scaling behavior. More interestingly,
the SSEC also exhibits scaling behavior, as manifested by the
collapse of the distributions in Fig. \ref{Fig:RI:PDF}. This
confirms that the scaling behavior of the return interval
distribution is significantly improved when the realized volatility
$R_2$ is adopted. The improved scaling behavior of the realized
volatility return interval distribution is consistent with the
scaling behavior of the return interval distribution for 4 Chinese
stocks reported by Qiu et al \cite{Qiu-Guo-Chen-2008-PA}\footnote{We
have discussed with Qiu and revealed that they defined the
volatility in a similar way as we did for the realized volatility
$R_2$. However, they did not clearly clarify this in their paper
\cite{Qiu-Guo-Chen-2008-PA}.}.

\begin{table}[htp]
 \centering
 \caption{Kolmogorov-Smirnov test of the return interval
distributions by comparing the statistic $KS$ with the critical
value $CV$ at the 5\% significance level for SSEC and $22$
constituent stocks.} \label{TB:KStest}
\begin{tabular}{cccccccc}
  \hline\hline
    \multirow{3}*[2mm]{Stock code} & \multicolumn{3}{c}{Volatility $R_1$} & & \multicolumn{3}{c}{Volatility $R_2$}\\  %
  \cline{2-4} \cline{6-8}
     & $KS$ & $CV$ & Scaling? & & $KS$ & $CV$ & Scaling?\\
  \hline
   SSEC   & $0.1170$ & $0.0434$ &  No & & $0.0455$ & $0.0502$ & Yes\\
  $600000$ & $0.0699$ & $0.0538$ &  No & & $0.0243$ & $0.0481$ & Yes\\
  $600019$ & $0.0223$ & $0.0672$ & Yes & & $0.0407$ & $0.0701$ & Yes\\
  $600026$ & $0.0277$ & $0.0511$ & Yes & & $0.0362$ & $0.0516$ & Yes\\
  $600028$ & $0.0670$ & $0.0653$ &  No & & $0.0374$ & $0.0652$ & Yes\\
  $600030$ & $0.0580$ & $0.0520$ &  No & & $0.0409$ & $0.0492$ & Yes\\
  $600036$ & $0.0436$ & $0.0576$ & Yes & & $0.0466$ & $0.0535$ & Yes\\
  $600073$ & $0.0501$ & $0.0500$ &  No & & $0.0345$ & $0.0452$ & Yes\\
  $600088$ & $0.0540$ & $0.0497$ &  No & & $0.0250$ & $0.0456$ & Yes\\
  $600100$ & $0.0462$ & $0.0500$ & Yes & & $0.0239$ & $0.0461$ & Yes\\
  $600104$ & $0.0149$ & $0.0573$ & Yes & & $0.0365$ & $0.0510$ & Yes\\
  $600110$ & $0.0911$ & $0.0506$ &  No & & $0.0958$ & $0.0457$ &  No\\
  $600171$ & $0.0377$ & $0.0530$ & Yes & & $0.0204$ & $0.0482$ & Yes\\
  $600320$ & $0.0515$ & $0.0506$ &  No & & $0.0305$ & $0.0469$ & Yes\\
  $600428$ & $0.0475$ & $0.0510$ & Yes & & $0.0345$ & $0.0501$ & Yes\\
  $600550$ & $0.0198$ & $0.0498$ & Yes & & $0.0198$ & $0.0467$ & Yes\\
  $600601$ & $0.0218$ & $0.0637$ & Yes & & $0.0460$ & $0.0569$ & Yes\\
  $600602$ & $0.0167$ & $0.0561$ & Yes & & $0.0316$ & $0.0513$ & Yes\\
  $600688$ & $0.0421$ & $0.0566$ & Yes & & $0.0391$ & $0.0531$ & Yes\\
  $600770$ & $0.0647$ & $0.0478$ &  No & & $0.0284$ & $0.0442$ & Yes\\
  $600797$ & $0.1179$ & $0.0647$ &  No & & $0.1220$ & $0.0568$ &  No\\
  $600832$ & $0.0610$ & $0.0474$ &  No & & $0.0203$ & $0.0469$ & Yes\\
  $600900$ & $0.0574$ & $0.0556$ &  No & & $0.0300$ & $0.0550$ & Yes\\
  \hline\hline
\end{tabular}
\end{table}

\subsection{Function form of $P_q(\tau)$}

Ren, Guo and Zhou have further performed the KS goodness-of-fit test
\cite{Clauset-Shalizi-Newman-2009-SIAMR,Gonzalez-Hidalgo-Barabasi-2008-Nature}
to study the particular form of the scaling function using the
volatility defined by $R_1$, and confirmed that the scaling function
of the stocks which show good scaling behavior could be approximated
by a stretched exponential form
\begin{equation}
f( \tau / \langle \tau \rangle )=c e^{- a (\tau / \langle \tau
\rangle) ^{\gamma}}. \label{Eq:pdf:scaling function}
\end{equation}
We adopt the same method to test the hypothesis that the empirical
distributions for different $q$ values for the realized volatility
defined by $R_2$ are coincident with a single stretched exponential
fit. Following Ren et al \cite{Ren-Guo-Zhou-2009-PA}, we test the
return interval distributions for two special $q$ values $q=2$ and
$5$ which behave most differently among all the distributions. Only
the 20 stocks which show scaling behaviors and the SSEC are tested.
If both return interval distributions for $q=2$ and $5$ are
identical to a {\em{same}} stretched exponential in the overlapping
region of the scaled return intervals, we can conclude that the
scaling function has a stretched exponential form.

In the case of the KS goodness-of-fit test, the $KS$ statistic
calculates the difference between the cumulative distribution $F_q$
of empirical return intervals and the cumulative distribution
$F_{\rm{SE}}$ from the fitted stretched exponential,
\begin{equation}
   KS = \max\left(|F_q-F_{\rm{SE}}|\right),~~~~q=2,5~.
   \label{Eq:KS2}
\end{equation}
A weighted $KS$ statistic, which is more sensitive on the edges of
the cumulative distribution, is defined as
\cite{Gonzalez-Hidalgo-Barabasi-2008-Nature}
\begin{equation}
KSW = \max
\left(\frac{|F_q-F_{\rm{SE}}|}{\sqrt{F_{\rm{SE}}(1-F_{\rm{SE}})}}
\right). \label{Eq:KSw}
\end{equation}
We generate 1000 synthetic samples from the best fitting
distribution, and calculate the $KS$ and $KSW$ statistics for the
synthetic data by taking the same measurements as we do for the
empirical data as
\begin{equation}
   KS_{\rm{sim}} = \max\left(|F_{\rm{sim}}-F_{\rm{sim,SE}}|\right)
   \label{Eq:KS2:sim}
\end{equation}
and
\begin{equation}
 KSW_{\rm{sim}} = \max
 \left(\frac{|F_{\rm{sim}}-F_{\rm{sim,SE}}|}{\sqrt{F_{\rm{sim,SE}}(1-F_{\rm{sim,SE}})}}\right),
 \label{Eq:KSw:sim}
\end{equation}
where $F_{\rm{sim}}$ is the CDF of each simulated synthetic sample
and $F_{\rm{sim,SE}}$ is the CDF of its best fit obtained from
integrating the fitted stretched exponential. The $p$-value is
defined as the frequency that $KS_{\rm{sim}}>KS$ or
$KSW_{\rm{sim}}>KSW$, and it can be regarded as the probability that
the empirical distribution is consistent with its best fit. The
tests are carried out for SSEC and $20$ constituent stocks, and the
resultant $p$-values are listed in Table
\ref{TB:goodness-of-fit-KS}.

\begin{table}[htp]
 \centering
 \caption{$KS$ and $KSW$ goodness-of-fit tests of the scaling function form of return interval
distributions for $q=2$ and $5$ by comparing empirical data with the
best stretched exponential fit and synthetic data with the best
stretched exponential fit. The stocks marked with $\star$ pass the
test using the $KS$ statistic, and the stocks marked with
$\star\star$ pass the test using both $KS$ and $KSW$ statistics
under the significant level of 5\%.} \label{TB:goodness-of-fit-KS}
\begin{tabular}{lccc|lccc|lccc}
  \hline\hline
   Code & $q$ & $p_{KS}$ & $p_{KSW}$ & Code & $q$ & $p_{KS}$ & $p_{KSW}$ & Code & $q$ & $p_{KS}$ & $p_{KSW}$\\
  \hline
  SSEC$^{\star}$   & $2$ & $0.321$ & $0.159$ & $600000^{\star\star}$ & $2$ & $0.338$ & $0.277$ & $600019^{\star}$ & $2$ & $0.074$ & $0.042$\\%
                   & $5$ & $0.203$ & $0.002$ &                       & $5$ & $0.434$ & $0.213$ &                  & $5$ & $0.530$ & $0.324$\\\hline%
  $600028$ & $2$ & $0.011$ & $0.001$ & $600030$ & $2$ & $0.168$ & $0.094$  & $600026^{\star\star}$ & $2$ & $0.630$ & $0.406$\\%
           & $5$ & $0.712$ & $0.841$ &          & $5$ & $0.018$ & $0.005$ &                       & $5$ & $0.288$ & $0.288$\\\hline%
  $600036^{\star\star}$ & $2$ & $0.707$ & $0.635$ & $600073^{\star\star}$ & $2$ & $0.417$ & $0.274$ & $600088^{\star\star}$ & $2$ & $0.443$ & $0.120$\\%
                        & $5$ & $0.209$ & $0.205$ &                       & $5$ & $0.411$ & $0.352$ &                       & $5$ & $0.066$ & $0.056$\\\hline%
  $600104$ & $2$ & $0.013$ & $0.006$ & $600100^{\star\star}$ & $2$ & $0.379$ & $0.236$ & $600171$ & $2$ & $0.016$ & $0.001$\\%
           & $5$ & $0.045$ & $0.017$ &                       & $5$ & $0.872$ & $0.665$ &          & $5$ & $0.005$ & $0$\\\hline%
  $600320^{\star\star}$ & $2$ & $0.744$ & $0.526$ & $600428$ & $2$ & $0.010$ & $0.006$ & $600550^{\star}$ & $2$ & $0.070$ & $0.042$\\%
                        & $5$ & $0.110$ & $0.051$ &          & $5$ & $0$     & $0.001$ &                  & $5$ & $0.564$ & $0.531$\\\hline%
  $600601^{\star\star}$ & $2$ & $0.493$ & $0.351$ & $600602^{\star}$ & $2$ & $0.522$ & $0.361$ & $600688$ & $2$ & $0.037$ & $0.001$ \\%
                        & $5$ & $0.417$ & $0.149$ &                  & $5$ & $0.130$ & $0.039$ &       & $5$ & $0.007$ & $0.002$\\\hline%
  $600770^{\star}$ & $2$ & $0.680$ & $0.441$ & $600832^{\star}$ & $2$ & $0.532$ & $0.246$ & $600900^{\star}$ & $2$ & $0.307$ & $0.256$\\%
                   & $5$ & $0.067$ & $0.015$ &                  & $5$ & $0.122$ & $0.006$ &                  & $5$ & $0.122$ & $0.005$\\%
  \hline\hline
\end{tabular}
\end{table}

By checking the $p$-values for both distributions for $q=2$ and $5$,
we can test the null hypothesis that the empirical PDFs can be
fitted well by a stretched exponential. Consider the significance
level of 1\%, if at least one $p$-value for $q=2$ or $5$ of an
individual stock is less than 1\%, then the null hypothesis is
rejected. Table \ref{TB:goodness-of-fit-KS} depicts the $p$-values
for both distributions for $q=2$ and $5$ for the volatility defined
by $R_2$. According to Table \ref{TB:goodness-of-fit-KS}, $17$
stocks as well as the SSEC pass the goodness-of-fit test using the
$KS$ statistic, and $12$ stocks pass the goodness-of-fit test using
the $KSW$ statistic. For the volatility defined by $R_1$, only $8$
stocks (out of the $20$ constituent stocks examined by the KS
goodness-of-fit test) pass the goodness-of-fit test using the $KS$
statistic and $7$ stocks pass the goodness-of-fit test using the
$KSW$ statistic \cite{Ren-Guo-Zhou-2009-PA}. Consider the
significance level of 5\%, $14$ stocks as well as the SSEC pass the
goodness-of-fit test using the $KS$ statistic, and $8$ stocks pass
the goodness-of-fit test using the $KSW$ statistic for the
volatility defined by $R_2$. In contrast, for the volatility defined
by $R_1$, $6$ stocks pass the goodness-of-fit test using the $KS$
statistic and $5$ stocks pass the goodness-of-fit test using the
$KSW$ statistic \cite{Ren-Guo-Zhou-2009-PA}. We find that the $KS$
and $KSW$ statistics provide very similar results for the volatility
defined by $R_1$, but the $KSW$ statistic is more sensitive than the
$KS$ statistic for the volatility defined by $R_2$. In principle,
the $p$-values of a stock are larger when the scaling of PDFs for
different $q$ is more significant.

\subsection{Dependence of $\gamma$ on $q$}

It has been shown that for the stocks which show multiscaling
behaviors, e.g. SSEC, the PDFs could also be well approximated by
the stretched exponential distribution but with different exponent
$\gamma$ for different threshold $q$ \cite{Ren-Zhou-2008-EPL}.
Therefore, we assume that the PDFs for the constituent stocks could
be approximated by the stretched exponential function, whether they
show good scaling behaviors or not. We fit the PDFs for different
threshold $q$ using a stretched exponential form
\begin{equation}
P_q(\tau)= b e^{- a (\tau / \langle \tau \rangle) ^{\gamma}}.
\label{Eq:RI:SE}
\end{equation}
where $a$, $b$ and $\gamma$ are dependent of $q$ if there is no
scaling. We investigate the relationship between the exponent
$\gamma$ and the threshold $q$ to further study the tendency of
return interval distribution with $q$.

\begin{figure}[h]
\centering
\includegraphics[width=6cm]{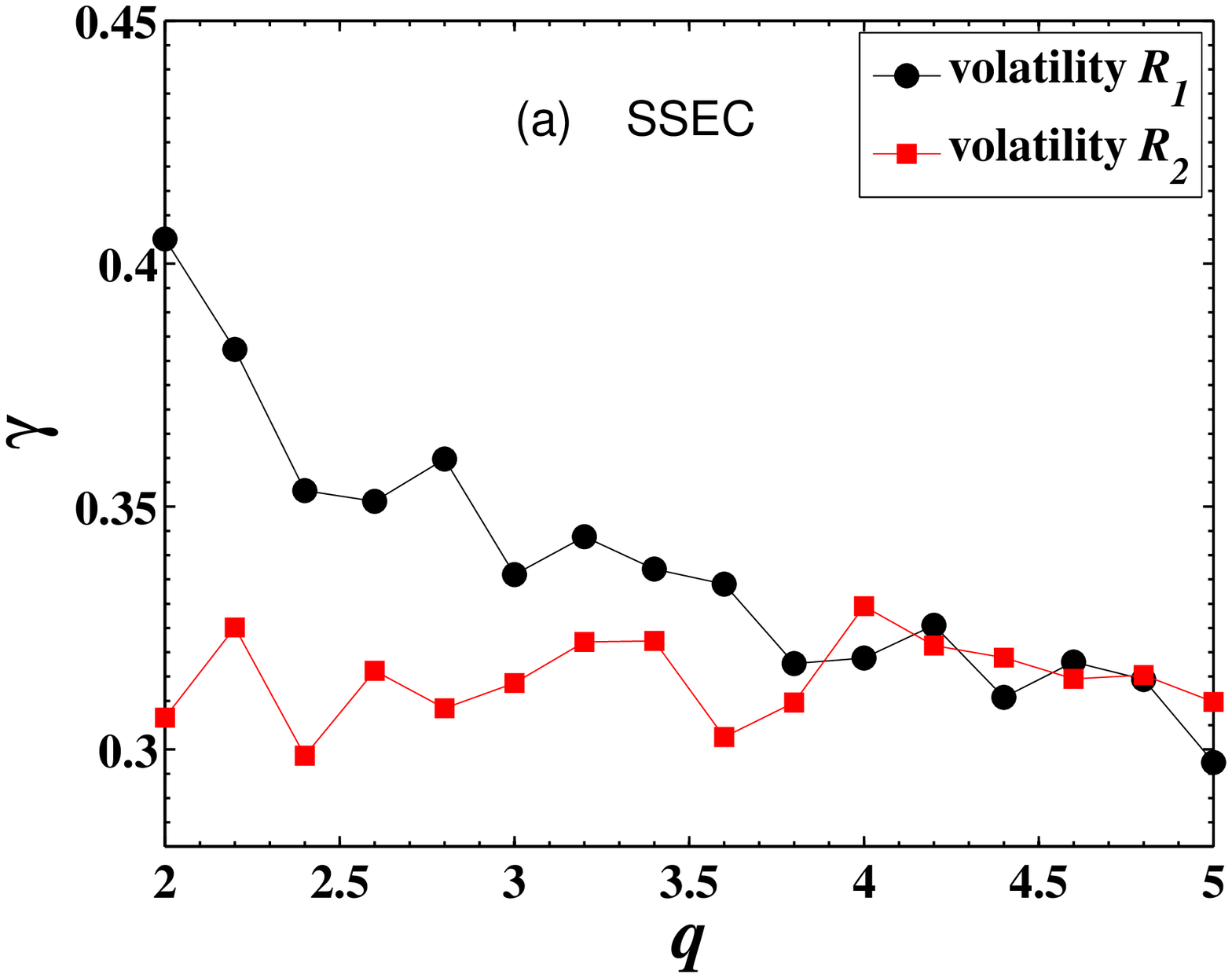}
\includegraphics[width=6cm]{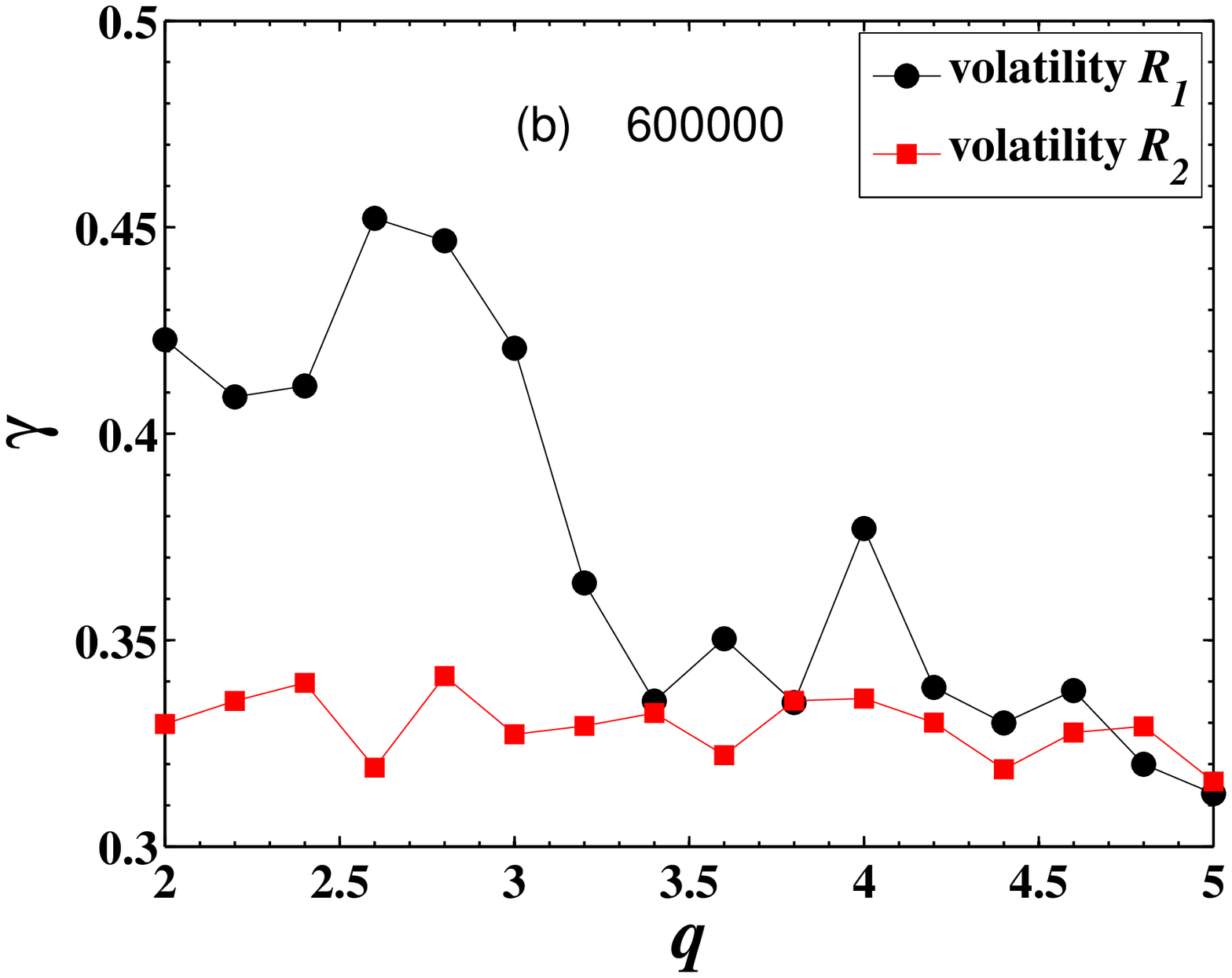}
\includegraphics[width=6cm]{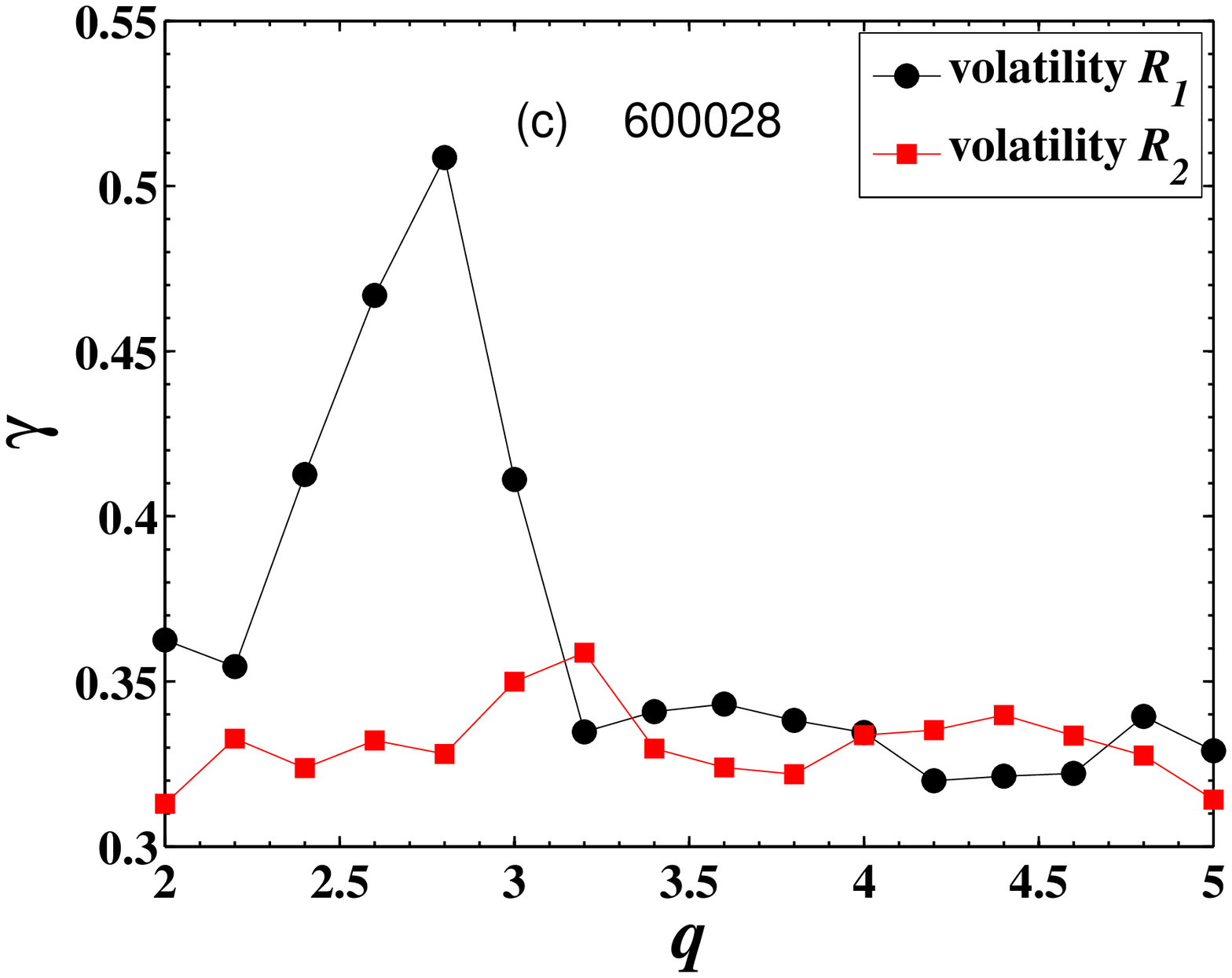}
\includegraphics[width=6cm]{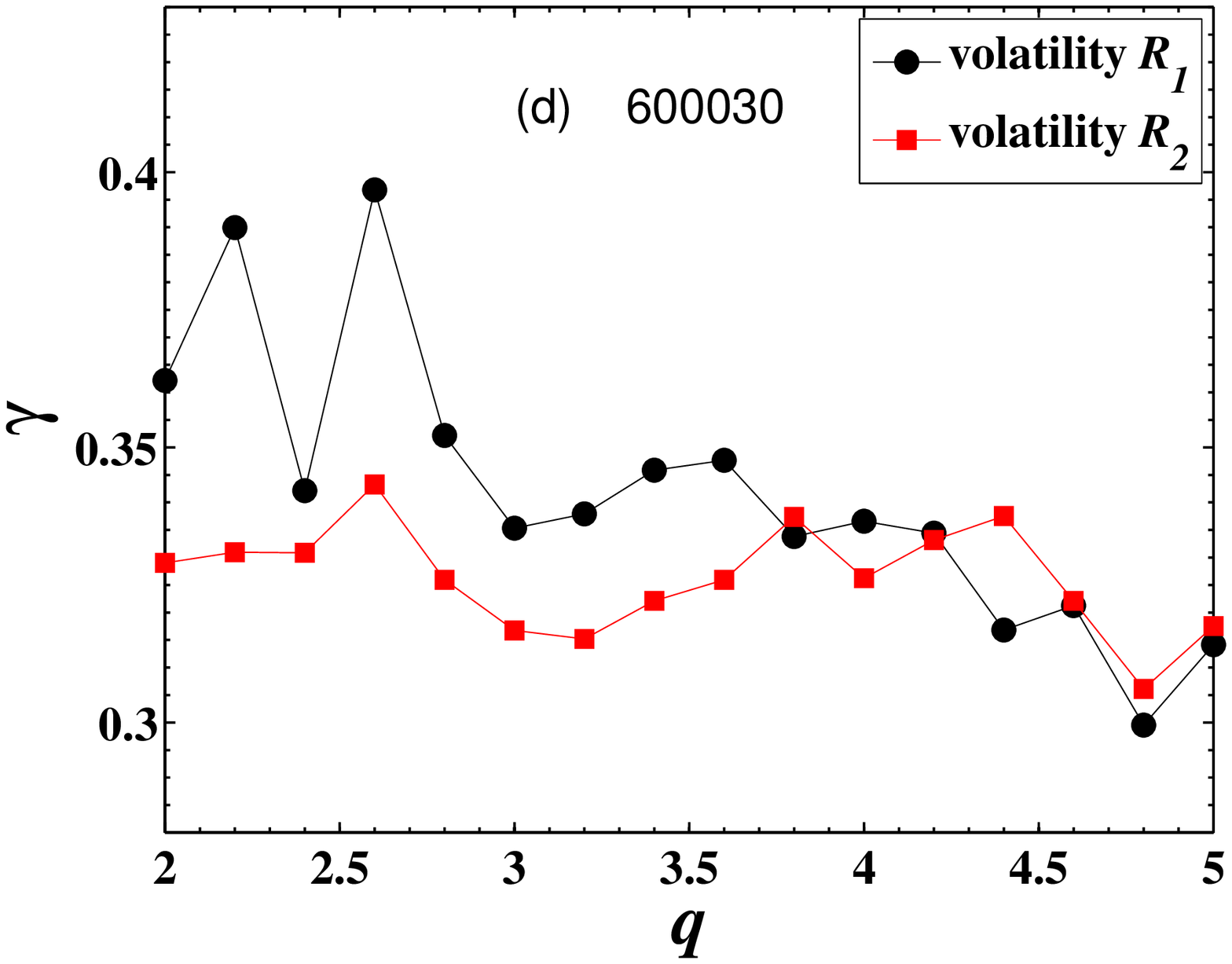}
\caption{(Color online) Exponent $\gamma$ vs. threshold $q$ for (a)
SSEC, (b) stock $600000$, (c) stock $600028$, and (d) stock
$600030$.} \label{Fig:gamma:q}
\end{figure}

In Fig. \ref{Fig:gamma:q} the exponents $\gamma$ are plotted as a
function of the threshold $q$ for SSEC and three representative
stocks $60000$, $600028$ and $600030$. Though the curve for the
volatility defined by $R_1$ fluctuates a little which may due to the
fitting errors caused by fluctuations, it shows an approximate
decreasing tendency with an increase of $q$. Regardless of the
fitting errors, the fitted exponent $\gamma$ mainly dominates the
shape of the return interval distribution. The decreasing tendency
of $\gamma$ further confirms our previous findings that the return
interval distributions for different $q$ values distinctly differ
from each other and show systematic deviation from scaling
\cite{Ren-Guo-Zhou-2009-PA}. Whereas for the volatility defined by
$R_2$, $\gamma$ slightly fluctuates but stays relatively constant in
comparison with that of the volatility defined by $R_1$. A similar
phenomenon is observed for other constituent stocks. This provides
additional evidence that the scaling behavior of the return interval
distribution for the realized volatility defined by $R_2$ is
significantly improved.

\subsection{Similarity of $P_q(\tau)$ for different stocks}

\begin{figure}[h]
\centering
\includegraphics[width=6cm]{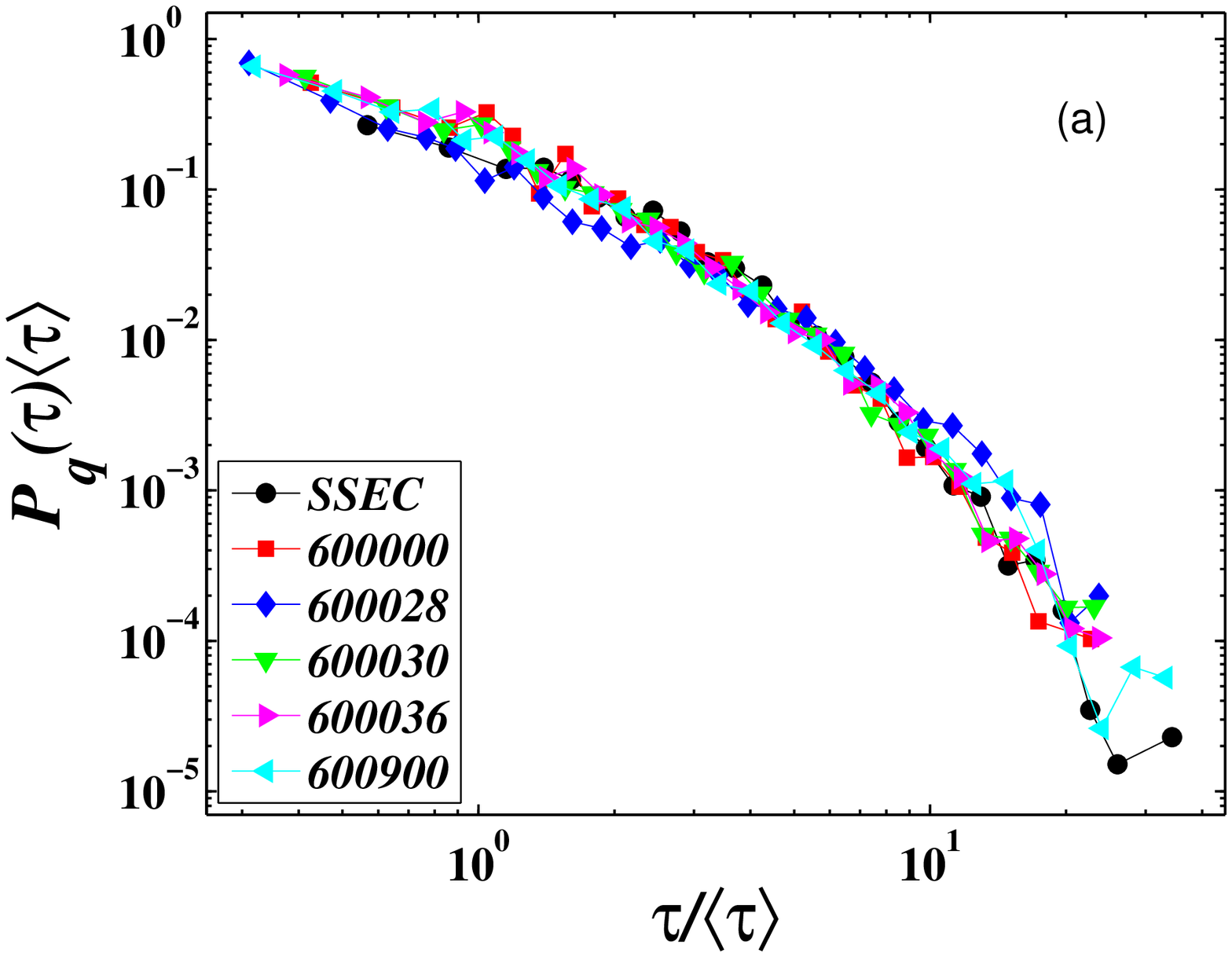}
\includegraphics[width=6cm]{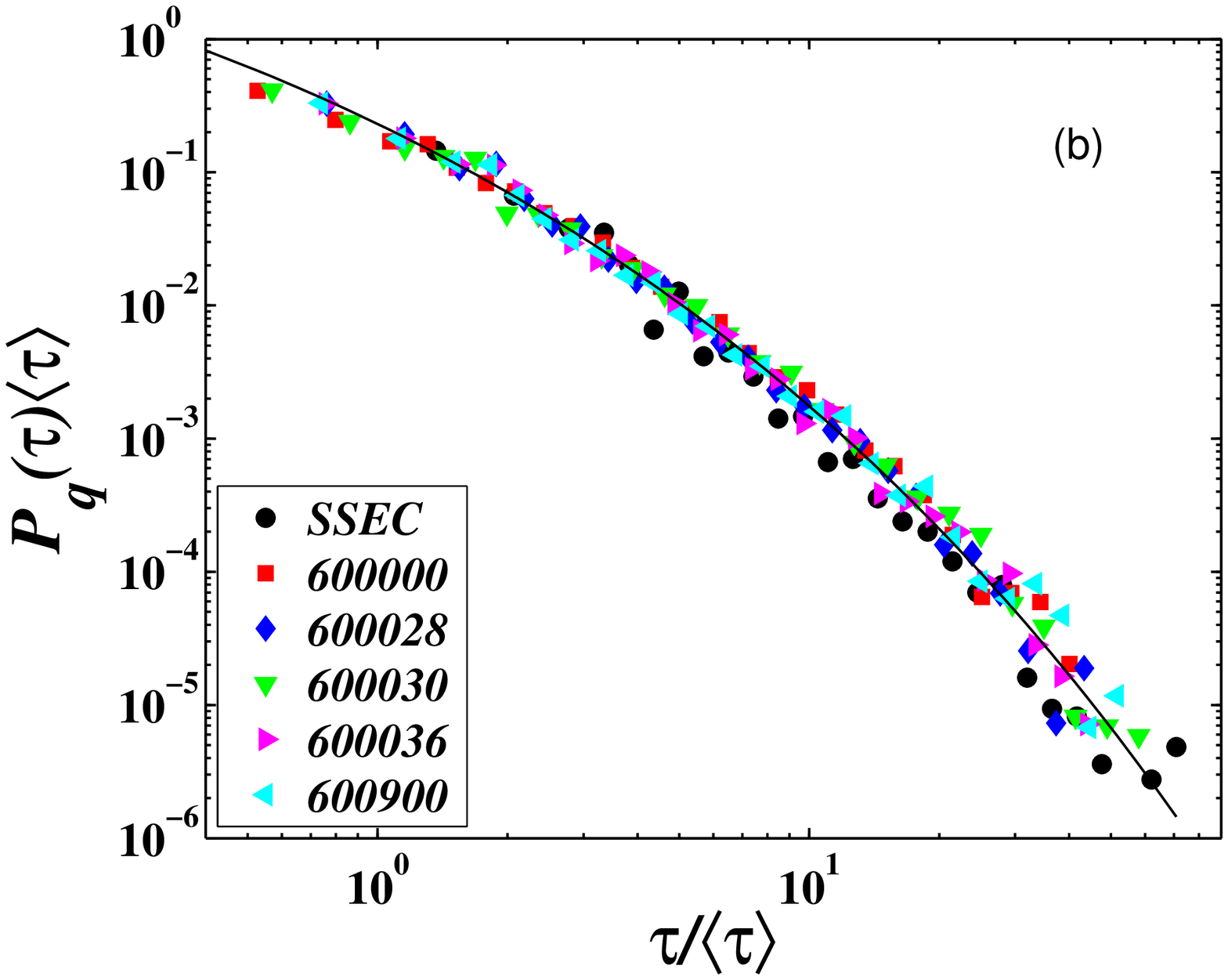}
\caption{(Color online) Probability distributions of the scaled
return intervals for SSEC and five representative constituent stocks
when $q=2$ for volatilities defined by (a) $R_1$ and (b) $R_2$. The
solid line in (b) is the stretched exponential fit with
$\gamma=0.26$.} \label{Fig:different:stocks}
\end{figure}

We also study the similarity of the return interval distributions
for different stocks. We fix the threshold $q=2$, and see how the
return interval distributions behave for different stocks. In Fig.
\ref{Fig:different:stocks} (a), the PDFs of the scaled return
intervals of the volatility defined by $R_1$ for SSEC and five
representative stocks are plotted. It is seen that the curves for
different stocks differ from each other and do not collapse to a
single curve. We then use the volatility defined by $R_2$ to
investigate the PDFs of the scaled return intervals for the SSEC and
the $5$ representative stocks. As shown in Fig.
\ref{Fig:different:stocks} (b), the curves for different stocks have
very similar shapes and seem to collapse on a single curve. This
indicates that the PDFs for different stocks may follow similar
scaling function if we use the volatility defined by $R_2$.

\section{Memory in the return intervals of realized volatility}
\label{S1:Memory}

\subsection{Short memory in realized volatility return intervals}

The memory effect is another important characteristic feature of the
stock markets. Empirical study has revealed that the memory effect
universally exists in the return intervals of various stock markets
for the volatility defined by $R_1$
\cite{Wang-Yamasaki-Havlin-Stanley-2006-PRE,Wang-Weber-Yamasaki-Havlin-Stanley-2007-EPJB,Jung-Wang-Havlin-Kaizoji-Moon-Stanley-2008-EPJB,Qiu-Guo-Chen-2008-PA,Ren-Guo-Zhou-2009-PA}.
We try to test if the memory effect retains in the return intervals
when we use the realized volatility defined by $R_2$.

\begin{figure}[h]
\centering
\includegraphics[width=6cm]{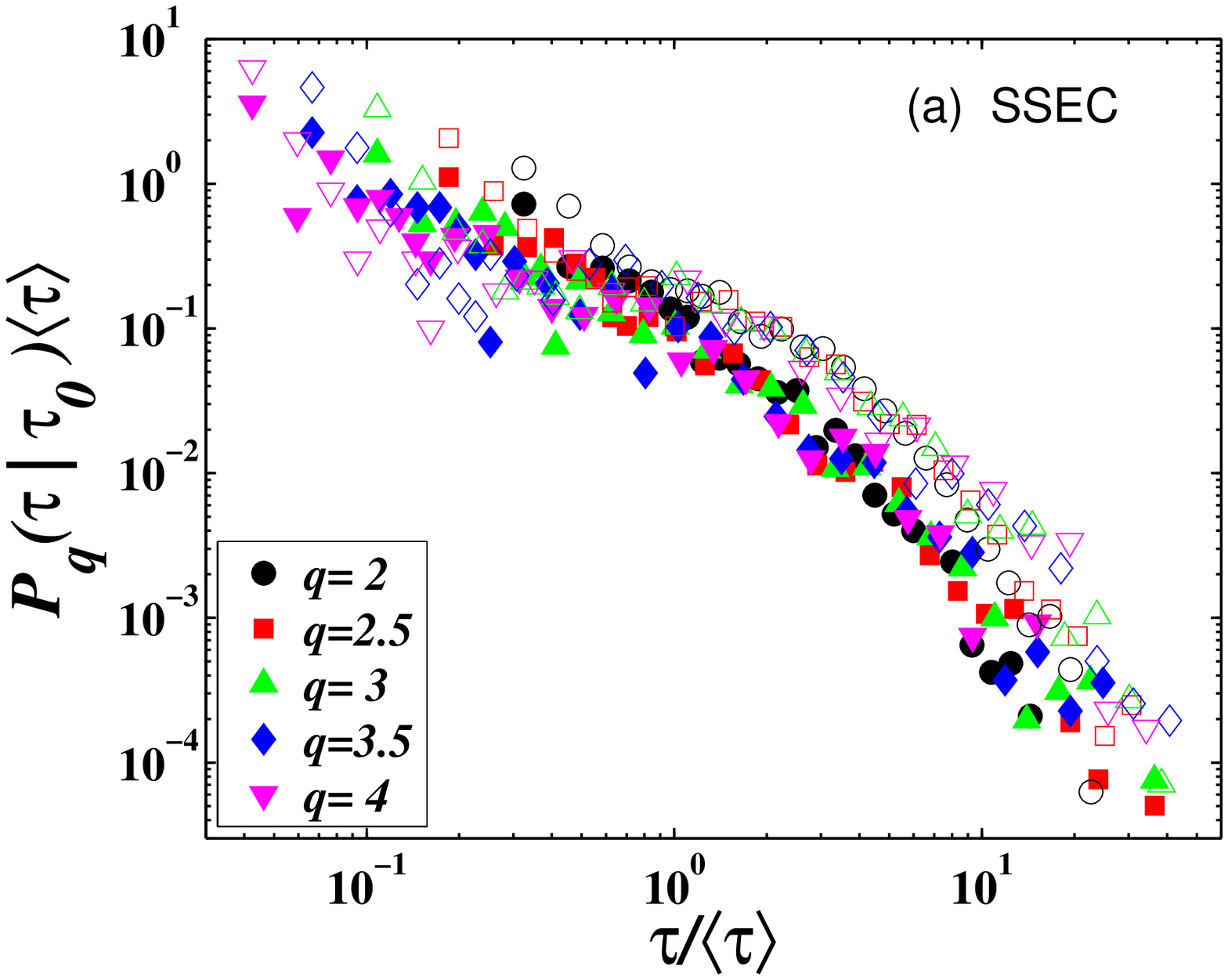}
\includegraphics[width=6cm]{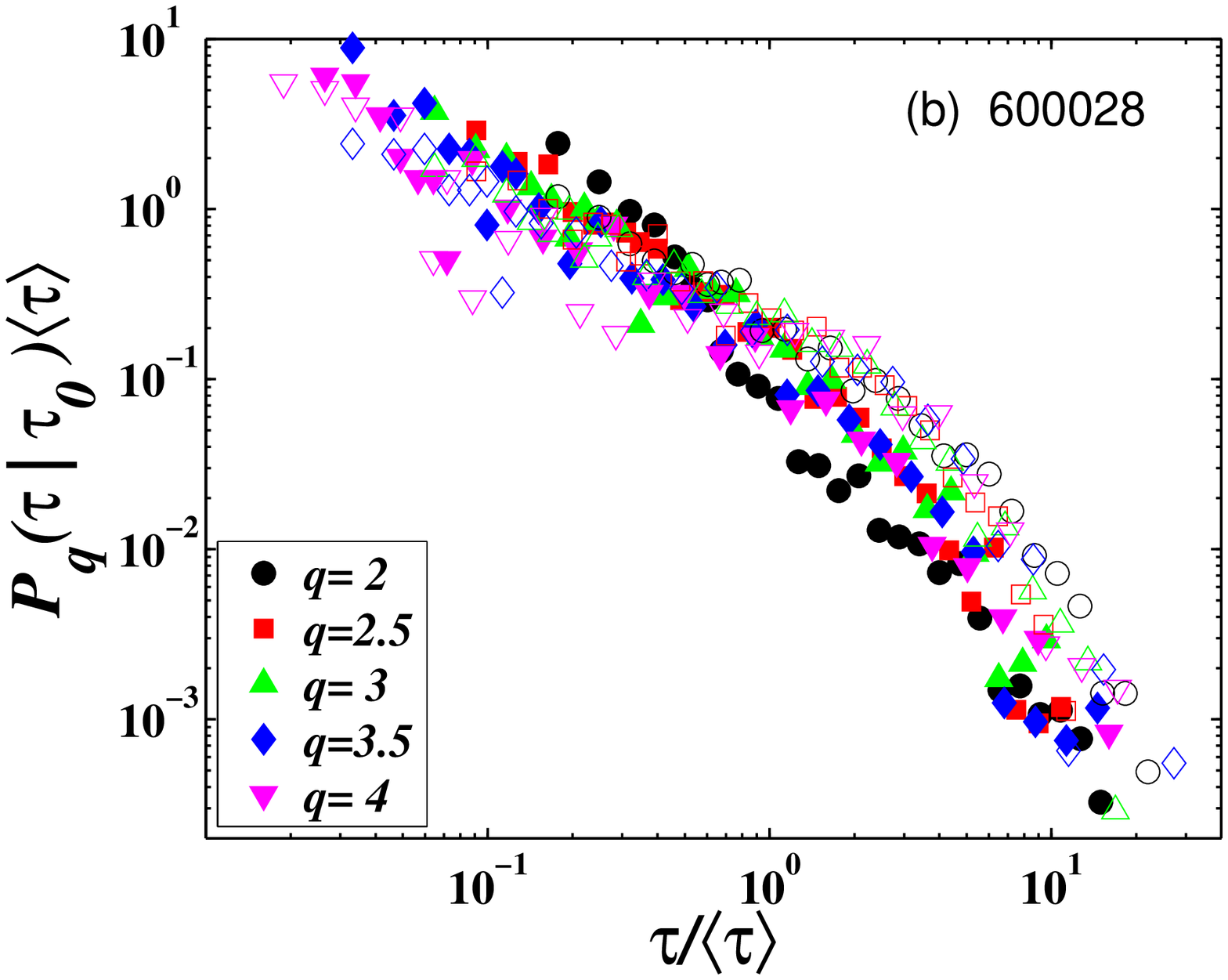}
\includegraphics[width=6cm]{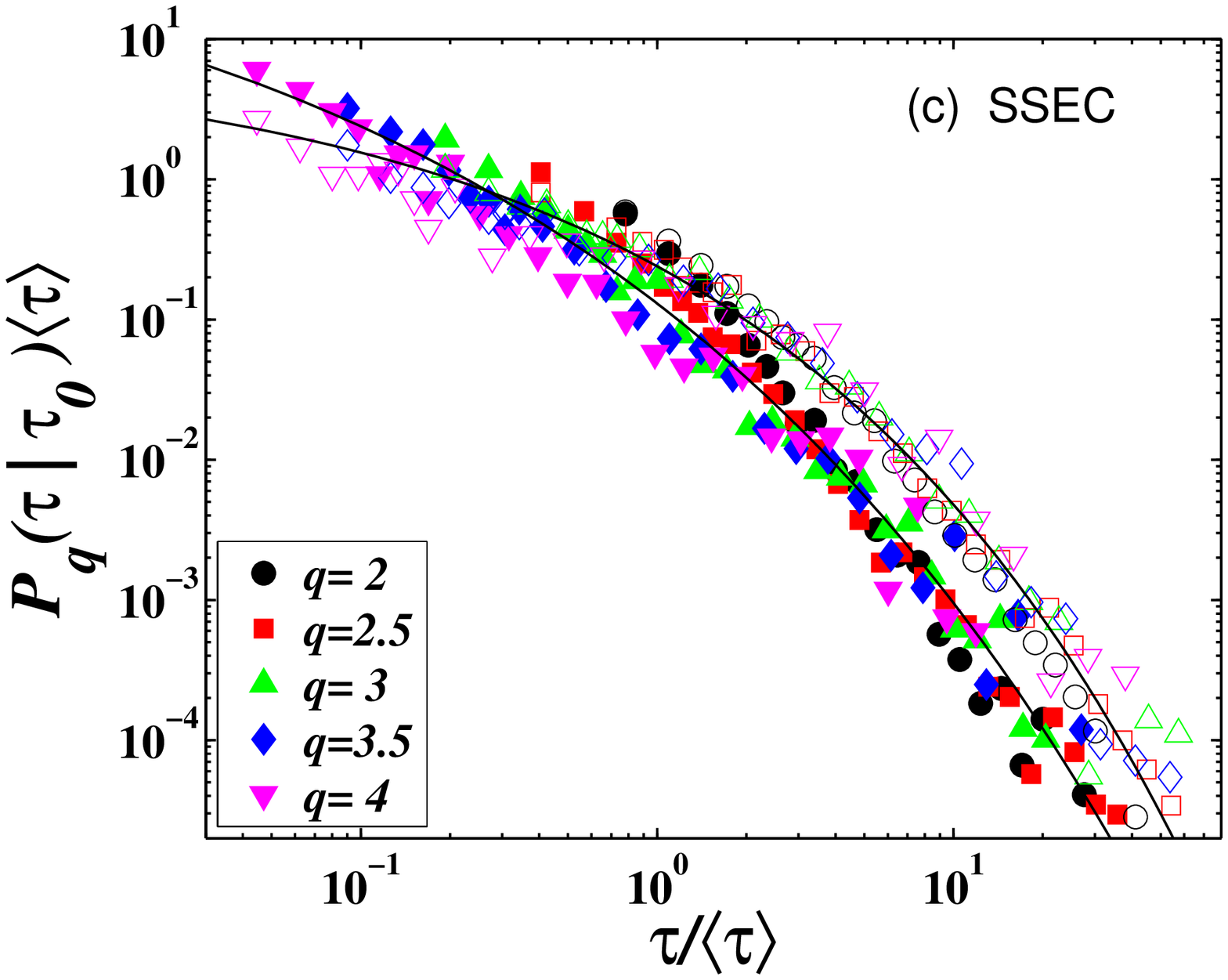}
\includegraphics[width=6cm]{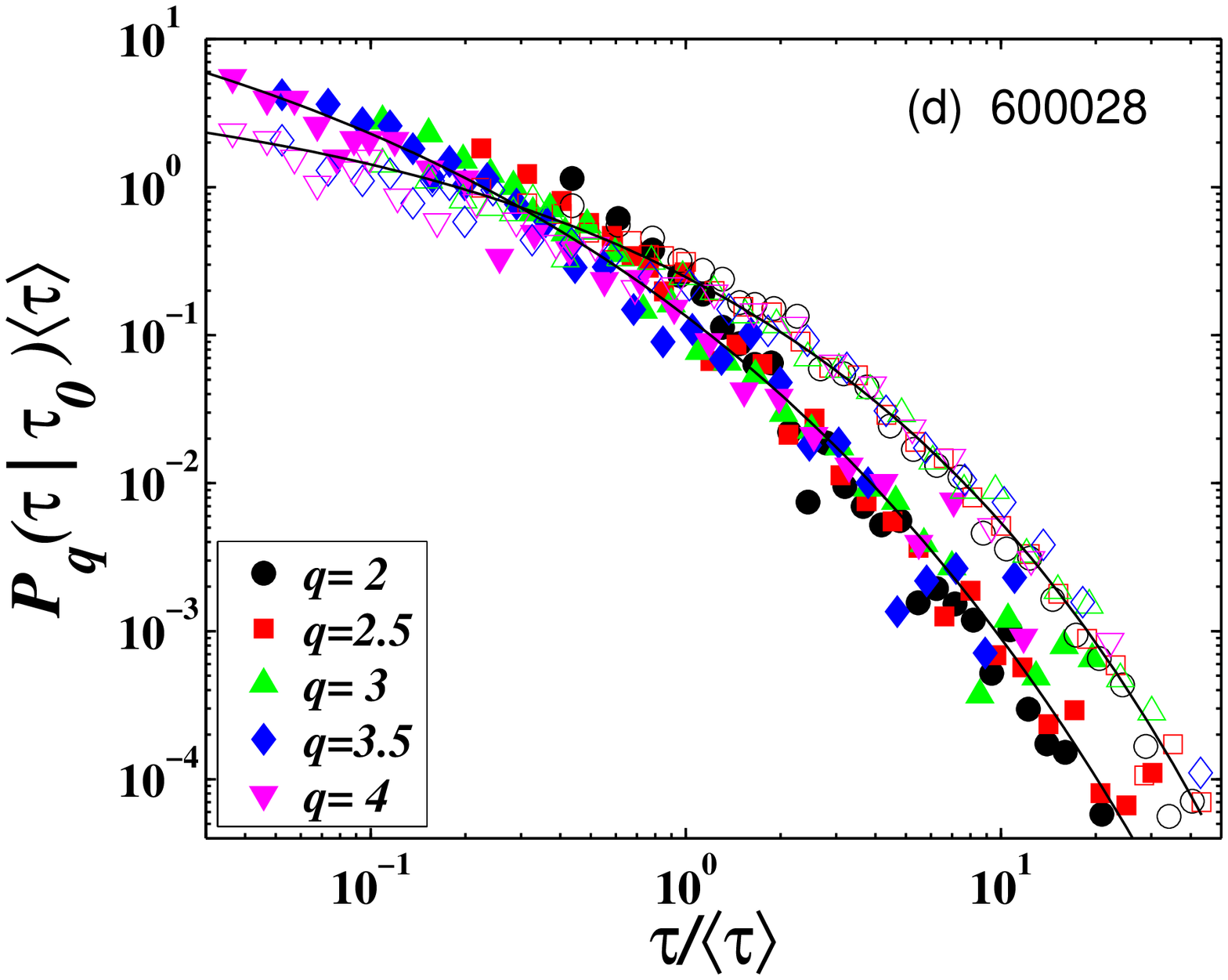}
\caption{(Color online) Conditional PDFs of the scaled return
intervals in the smallest $1/4$ subset (filled symbols) and the
largest $1/4$ subset (open symbols) for volatilities defined by (a)
$R_1$ for SSEC, (b) $R_1$ for stock $600028$, (c) $R_2$ for SSEC,
and (d) $R_2$ for stock $600028$. The Solid lines in (c) and (d) are
the stretched exponential fits.} \label{Fig:condictional PDF}
\end{figure}

We first investigate the short-term memory of the return intervals
by calculating the conditional PDF $P_q(\tau|\tau_0)$, defined as
the probability to find an interval $\tau$ immediately after an
interval $\tau_0$. To get better statistics, we study
$P_q(\tau|\tau_0)$ for a bin of $\tau_0$. The entire interval
sequences are arranged in an ascending order and partitioned into
four bins with equal size. $P_q(\tau|\tau_0)$ for $\tau_0$ in the
smallest and biggest quarter bins for SSEC and stock $600028$ with
volatility defined by $R_1$ are illustrated in Fig.
\ref{Fig:condictional PDF} (a) and (b). The curves intersperse with
each other for small scaled return intervals, and it is hard to
distinguish the curve for $\tau_0$ in the smallest subset from that
for $\tau_0$ in the biggest subset. In comparison with that of the
volatility defined by $R_1$, $P_q(\tau|\tau_0)$ for the same stock
index and the individual stock for the volatility defined by $R_2$
are plotted in Fig. \ref{Fig:condictional PDF} (c) and (d). One
observes that the scaling behavior of $P_q(\tau|\tau_0)$ is
significantly improved for the volatility defined by $R_2$
especially for small scaled return intervals: the curves for all
thresholds for $\tau_0$ in the smallest subset and biggest subset
approximately collapse onto two separate solid curves.
$P_q(\tau|\tau_0)$ shows bigger probabilities for $\tau_0$ in the
smallest (biggest) subset when $\tau/\langle \tau \rangle$ is small
(big), and this indicates that small (big) intervals $\tau_0$ tend
to be followed by small (big) intervals $\tau$. This can be regarded
as a proof of the existence of short-term memory in realized
volatility return intervals, while all the $P_q(\tau|\tau_0)$ curves
for the shuffled data collapse to a single exponential curve (not
shown).

\subsection{Long memory in realized volatility and its return
intervals}

To further investigate the long-term memory of the realized
volatility and its return intervals we use the detrended fluctuation
analysis (DFA) method
\cite{Peng-Buldyrev-Havlin-Simons-Stanley-Goldberger-1994-PRE,Hu-Ivanov-Chen-Carpena-Stanley-2001-PRE,Kantelhardt-Bunde-Rego-Havlin-Bunde-2001-PA,Chen-Ivanov-Hu-Stanley-2002-PRE,Chen-Hu-Carpena-Bernaola-Galvan-Stanley-Ivanov-2005-PRE},
known as a general method of examining the long-term correlation in
time series analysis. The DFA method computes the average
fluctuation $F(l)$ of the cumulative series $y(t)=\sum_{t'=1}^{t}
x(t')$ of data series $x(t')$ as
\begin{equation}
   F(l) =\frac{1}{N_l} \sum_{i}^{N_l} \sum_{t=1}^{l}
   (y(t)-\tilde{y}(t))^2,
   \label{Eq:DFA}
\end{equation}
where $N_l$ is the number of windows with fixed $l$ data points, and
$\tilde{y}(t)$ is a local linear estimation for $y(t)$ in a certain
window $i$. It is expected that $F(l)$ scales with $l$ as
\begin{equation}
   F(l)\sim l^ \alpha,
\end{equation}
The DFA method provides an accurate estimation of long-range
correlation which do not depend on the length of the time series,
and the scaling exponent $\alpha$ is supposed to be equal to the
Hurst exponent when $\alpha \leq 1$
\cite{Coronado-Carpena-2005-JBP}. Generally, for $\alpha>0.5$ the
time series are long-term correlated, and for $\alpha=0.5$ the time
series are uncorrelated.

\begin{figure}[h]
\centering
\includegraphics[width=6cm]{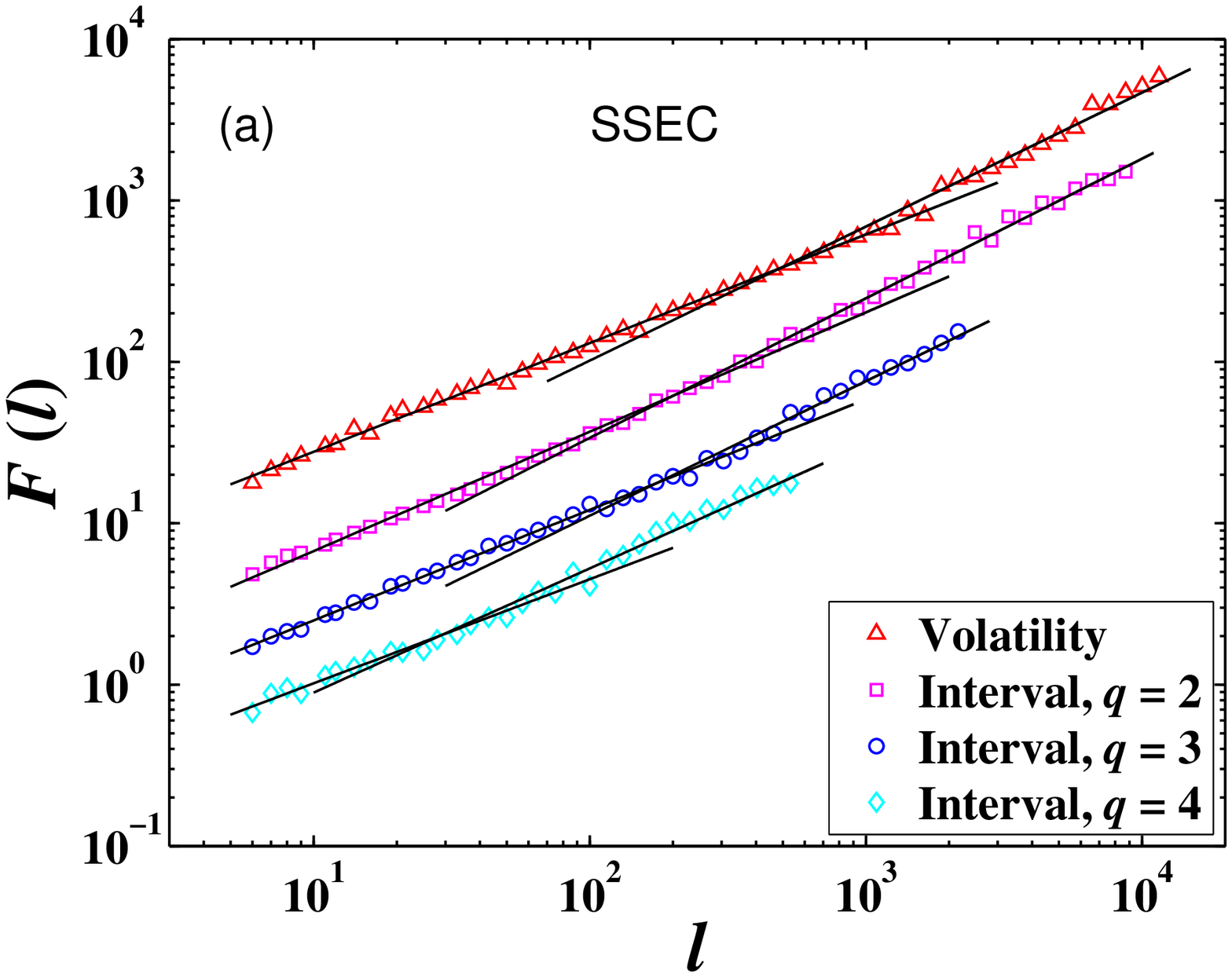}
\includegraphics[width=6cm]{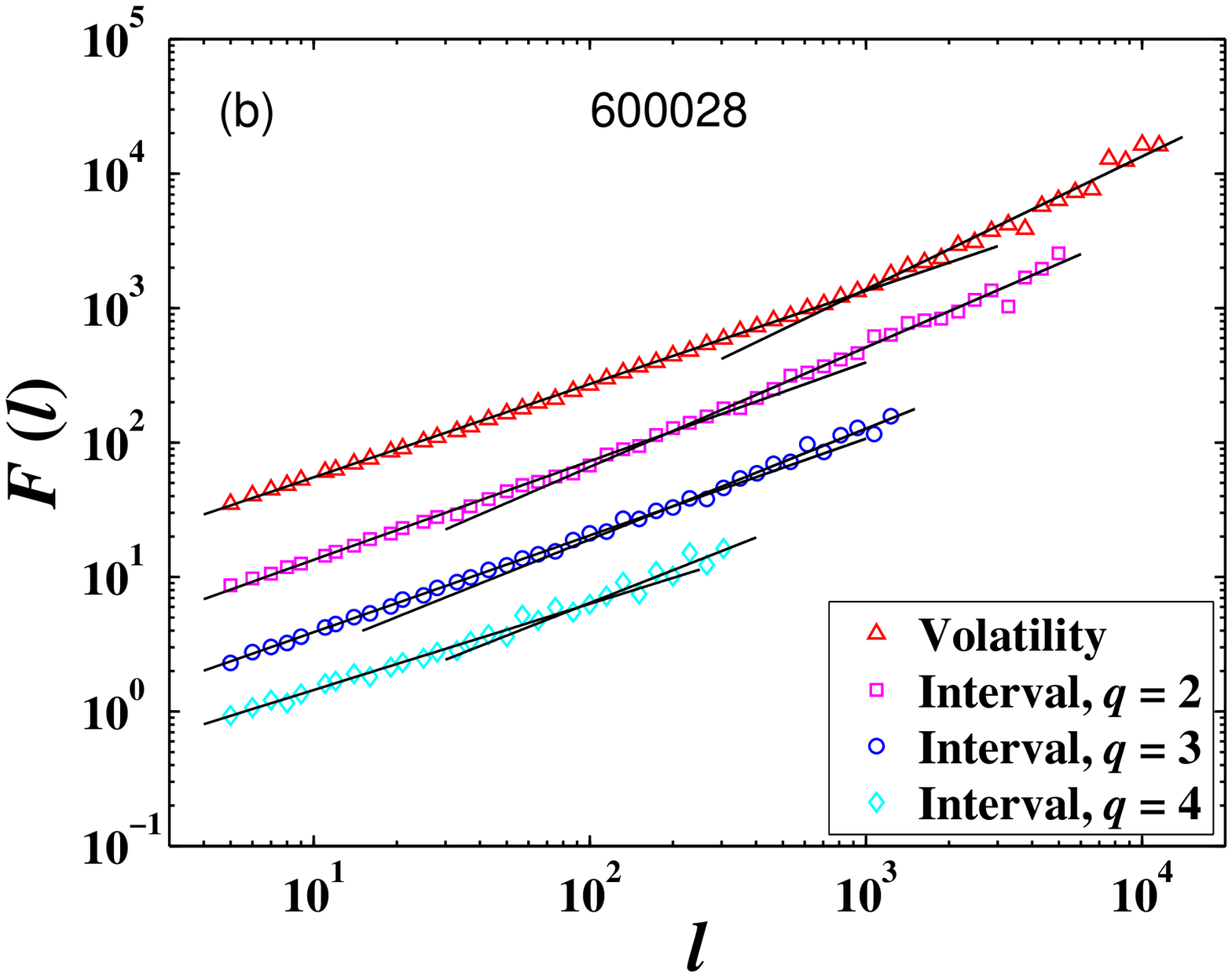}
\caption{(Color online) Detrended fluctuation functions $F(l)$ of
the realized volatility return intervals and the original realized
volatilities for (a) SSEC and (b) stock 600028. The curves are
vertically shifted for clarity.} \label{Fig:DFA}
\end{figure}

Fig. \ref{Fig:DFA} presents the detrended fluctuation functions
$F(l)$ of the realized volatility for SSEC and stock $600028$. A
crossover behavior is observed: for small scales of $l$, $F(l)$
obeys a power law with a relatively small exponent; while for large
scales of $l$, $F(l)$ obeys a power law with a relatively large
exponent. The exponent $\alpha$ is estimated to be $0.68\pm0.01$ and
$0.86\pm0.02$ for SSEC and $0.69\pm0.01$ and $0.98\pm0.03$ for stock
$600028$ respectively in small scale and large scale regions,
apparently larger than $0.5$. Similar results are observed for other
constituent stocks. Therefore, we can conclude that the realized
volatilities are long-term correlated.

\begin{figure}[h]
\centering
\includegraphics[width=5.2cm]{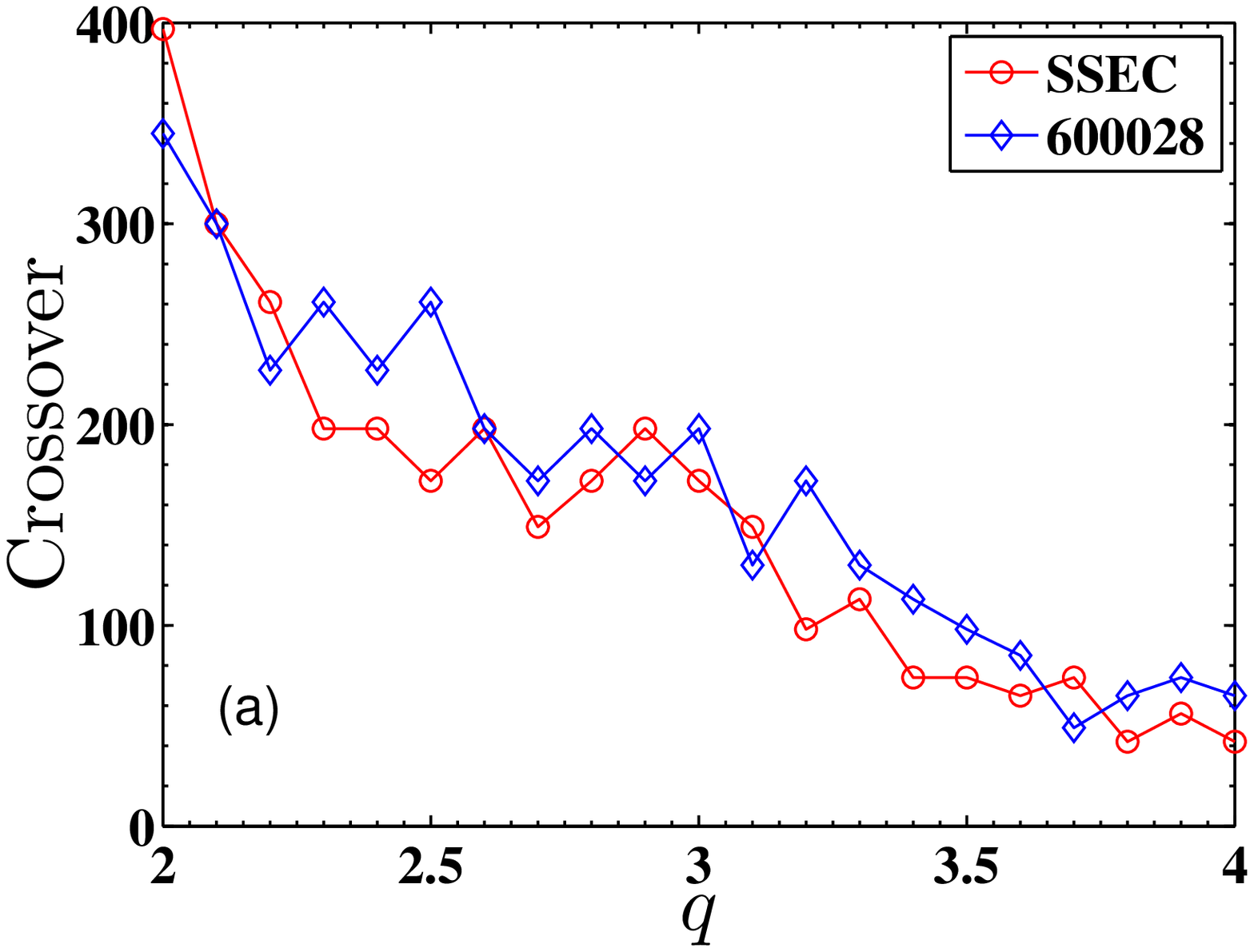}
\includegraphics[width=5.2cm]{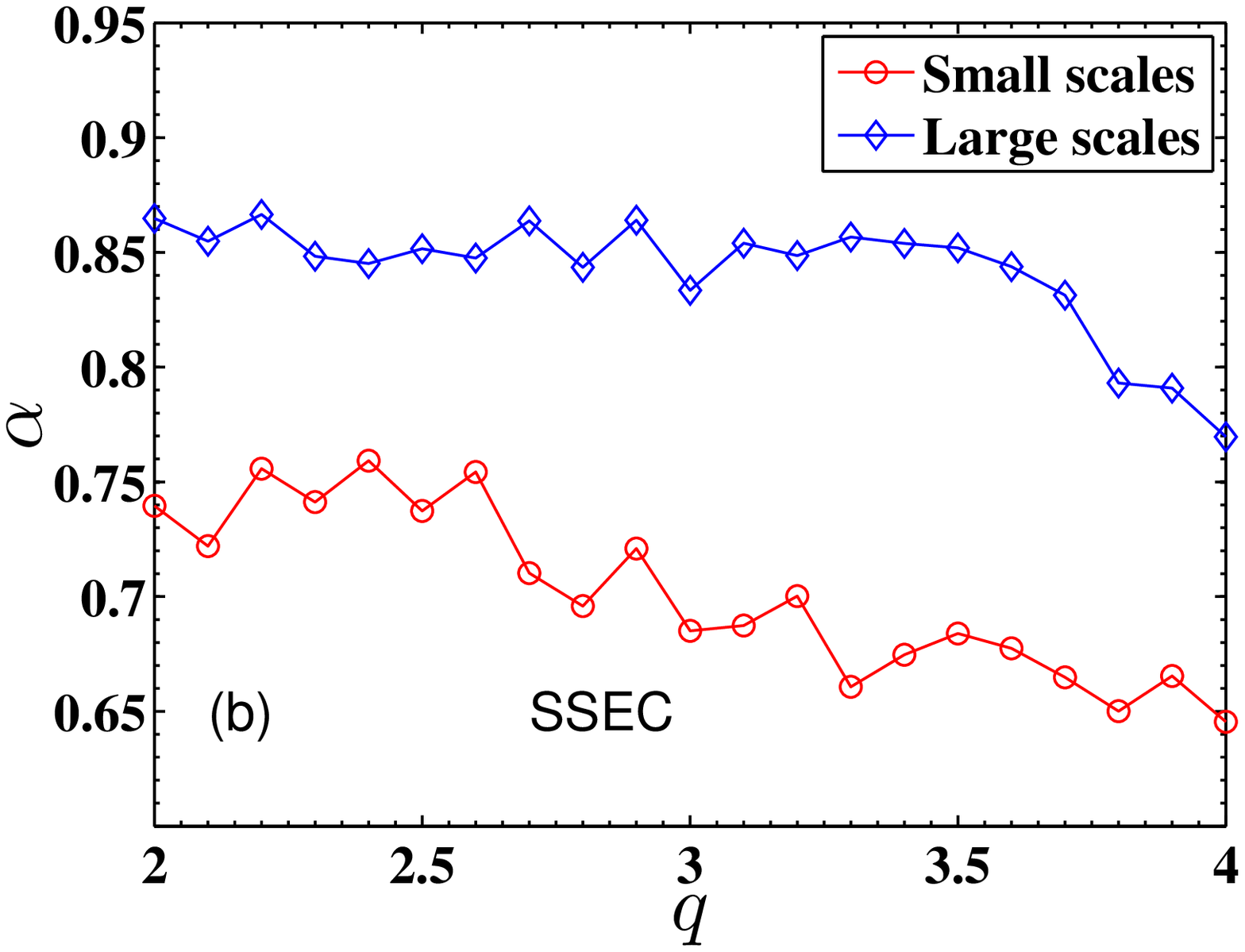}
\includegraphics[width=5.2cm]{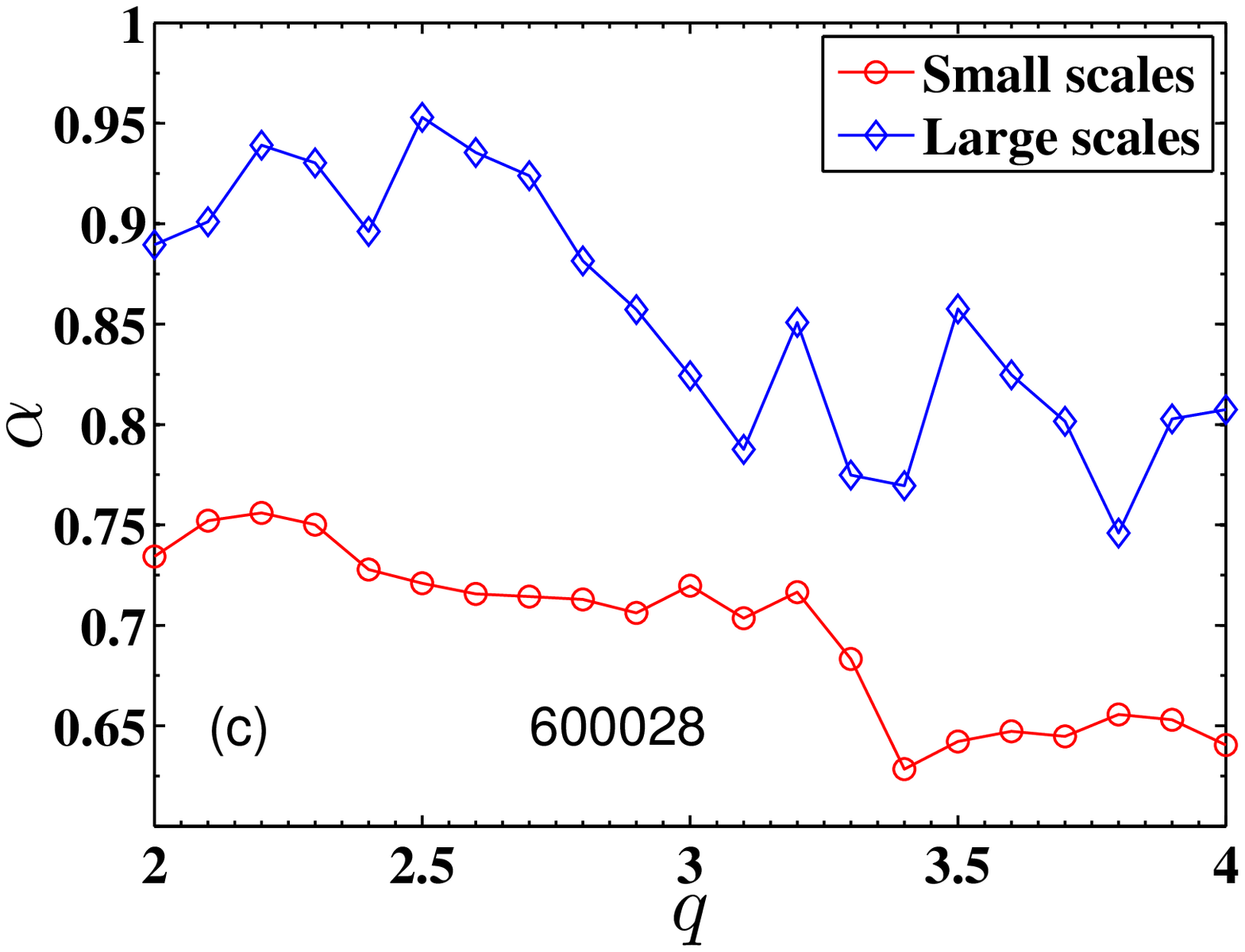}
\caption{(Color online) (a) Crossover point as a function of $q$ for
SSEC (circles) and stock 600028 (diamonds). Exponent $\alpha$ of the
realized volatility return intervals for large scales (diamonds) and
small scales (circles) for (b) SSEC and (c) stock 600028.}
\label{Fig:Hurst}
\end{figure}

We also compute the detrended fluctuation functions $F(l)$ of the
return intervals of realized volatility as shown in
Fig.~\ref{Fig:DFA}, and observe similar crossover behavior. The
estimation of the crossover point which separates the two power-law
regions becomes an important task, since it essentially affects the
determination of exponent $\alpha$. A simple least squares
estimation method is applied to determine the value of the threshold
by minimizing the square distance between $F(l)$ and its best
power-law fits in small scale and large scale regions. In fact we
have used the same least squares estimation method to find out the
crossover point for $F(l)$ of the realized volatility. In Fig.
\ref{Fig:DFA} (a) and (b), the solid lines are power-law fits in
small scale and large scale regions, respectively. Apparently the
crossover tends to appear at smaller scales when the threshold $q$
increases as quantitatively illustrated in Fig. \ref{Fig:Hurst} (a).

As with this least squares estimation method, we can further test
the relation between the exponent $\alpha$ of the return intervals
and the threshold $q$, and see how the long-term memory of the
return intervals varies with the change of $q$. In Fig.
\ref{Fig:Hurst} (b) and (c), the exponent $\alpha$ for SSEC and
stock $600028$ are plotted as a function of the threshold $q$. The
curves fluctuate a little for relatively large $q$ due to the poor
statistics of reduced interval samples. In general, the Hurst
exponents in both regions for small scales and large scales show
decreasing tendencies when $q$ increases. Though the long-term
correlation of the return intervals is weakened with the increase of
$q$, the exponent $\alpha$ for all thresholds is apparently larger
than $0.5$. For the shuffled realized volatility data, the exponent
$\alpha$ of the return intervals displays a value close to $0.5$.
This indicates that the long-term memory of the return intervals may
arise from the long-term memory of original volatility records.

\section{Conclusion}
\label{S1:Conclusion}

In summary, we have studied the statistical properties of the return
intervals of 1-min realized volatility based on the high-frequency
intraday data for the SSEC index and $22$ liquid constituent stocks.
The Kolmogorov-Smirnov test shows that $20$ stocks (out of $22$
stocks) and the SSEC exhibit scaling behaviors. We found that the
scaling behavior of the return interval distribution of the realized
volatility is significantly improved compared with that of the
ordinary volatility defined by the closest tick prices to the minute
marks. We further adopted the KS goodness-of-fit test using the $KS$
and weighted $KSW$ statistics to study the particular form of the
scaling distribution, and found the scaling function for 8
constituent stocks can be well-approximated by a stretched
exponential distribution $f ( \tau / \langle \tau \rangle )=c e^{- a
(\tau / \langle \tau \rangle) ^{\gamma}}$. We calculated the
relation between the exponent $\gamma$ estimated from the stretched
exponential fit of $P_q (\tau)$ and the threshold $q$, and further
demonstrated the improved scaling behavior of the realized
volatility. The similarity of $P_q (\tau)$ for different stocks is
also observed for the realized volatility.

We then investigated the memory effect of the realized volatility
return intervals for the SSEC and $22$ constituent stocks.
Short-term memory is revealed by the observation of the conditional
probability distribution $P_q(\tau|\tau_0)$ which also shows good
scaling behavior for the realized volatility. Using the DFA method,
we found that long-term memory exists in both realized volatility
and its return intervals, and the exponent $\alpha$ of the realized
volatility return intervals shows a decreasing tendency with the
increase of the threshold $q$.

\bigskip
{\textbf{Acknowledgments:}}

This work was partially supported by the Shanghai Educational
Development Foundation (2008CG37 and 2008SG29), the National Natural
Science Foundation of China (70501011), and the Program for New
Century Excellent Talents in University (NCET-07-0288).

%\pagebreak
\bibliography{E:/papers/Auxiliary/Bibliography}
%\bibliography{E:/paper/bibfile/Bibliography}

\end{document}